\documentclass[a4paper,11pt]{article}
\usepackage[utf8]{inputenc}
\pdfoutput=1
\usepackage{jheppub,float}
\usepackage[normalem]{ulem}
\usepackage{xcolor}
\usepackage{amsmath}
\usepackage{framed}
\usepackage{mathrsfs,stackrel, amssymb,amsthm,tikz,graphicx,accents}
\usepackage{slashed,ccaption}
\usepackage{mathrsfs, calligra}
\usepackage{multirow}
\usepackage{leftidx}
\usepackage{import}
\usepackage{multirow}
\usepackage{amsfonts}
\colorlet{shadecolor}{lightgray}
\usepackage{hyperref}
\usepackage{parskip}

\newcommand{\cR}{\mathcal{R}}
\newcommand{\es}{\texttt{s}}

\newcommand{\cQ}{\mathcal{Q}}

\newcommand{\BMSt}{\mathfrak{bms}_{3}}
\newcommand{\BMSf}{\mathfrak{bms}_{4}}

\newcommand{\half}{+\frac{1}{2}}
\newcommand{\mhalf}{-\frac{1}{2}}

\definecolor{darkgreen}{RGB}{27,130,45}
\definecolor{darkpurple}{RGB}{108,0,217}

\newcommand\anote[1]{\textcolor{magenta}{\bf [Arpita:\,#1]}}

\newcommand\dnote[1]{\textcolor{orange}{\bf [Debangshu:\,#1]}}
\newcommand\hnote[1]{\textcolor{darkgreen}{\bf [Hamid:\,#1]}}

\title{Supersymmetrization of deformed BMS algebras}

\author[a]{Nabamita Banerjee,}
\author[a,b]{Arpita Mitra,}
\author[c]{Debangshu Mukherjee,}
\author[d]{H. R. Safari,}

\affiliation[a]{Indian Institute of Science Education \& Research Bhopal,\\
	Bhopal Bypass Road, Bhauri, Bhopal 420 066, India}
\affiliation[b]{Physics Division, National Center for Theoretical Sciences, Taipei 10617, Taiwan}
\affiliation[c]{Indian Institute of Science Education \& Research Thiruvananthapuram,\\
	Vithura 695551, Kerala, India}
\affiliation[d]{School of Physics, Institute for Research in Fundamental Sciences (IPM),\\P.O.Box 19395-5531, Tehran, Iran}
\emailAdd{nabamita@iiserb.ac.in, arpitam@iiserb.ac.in, debanshu@iisertvm.ac.in, hrsafari@ipm.ir}
\abstract{\texorpdfstring{$W(a,b)$}{W(a,b)} and \texorpdfstring{$W(a,b;\bar{a},\bar{b})$ algebras}{W(a,b;bar{a},bar{b})} are deformations of $\BMSt$ and $\BMSf$ algebra respectively. We present an $\mathcal{N}=2$ supersymmetric extension of \texorpdfstring{$W(a,b)$}{W(a,b)} and \texorpdfstring{$W(a,b;\bar{a},\bar{b})$ algebra}{W(a,b;bar{a},bar{b})} in presence of $R-$symmetry generators that rotate the two supercharges. For  \texorpdfstring{$W(a,b)$}{W(a,b)} our construction includes most generic central extensions of the algebra. In particular we find that $\mathcal{N}=2$ $\BMSt$ algebra  admits a new central extension that has so far not been reported in the literature. 
For \texorpdfstring{$W(a,b;\bar{a},\bar{b})$}{W(a,b;bar{a},bar{b})}, we find that an infinite $U(1)_V \times U(1)_A$ extension of the algebra is not possible with linear and quadratic  structure constants for generic values of the deformation parameters. This implies a similar constraint for $U(1)_V \times U(1)_A$ extension of $\mathcal{N}=2 \, \BMSf$ algebra. }

\dedicated{Last Updated: \today}

\begin{document} 
\begin{flushright}
  {\small 
   IPM/P-2022/01
  }
\end{flushright}
\maketitle
\flushbottom

\section{Introduction and Summary}
For any theory its asymptotic symmetries are of immense physical significance. The symmetries at the asymptotic boundary of a theory depends on the boundary fall off of its constituents fields. In most examples, the asymptotic symmetry is usually enlarged compared to the bulk symmetry of the theory. However, the bulk symmetry must be contained in the asymptotic symmetry group algebra as a subalgebra. For a theory in asymptotically flat spacetimes, if one recedes from sources towards null infinity, at any finite radial distance from the source one expects the symmetry algebra to be just Poincar\'{e}. However at null infinity, the asymptotic symmetry algebra in the Bondi gauge is enhanced to the Bondi-Metzner-Sachs or the $\mathfrak{bms}$ algebra generated by infinite number of generators known as supertranslations \cite{BMvdB:1962, Sachs:1962}. One can further extend the $\mathfrak{bms}$ algebra by including superrotations \cite{Barnich:2009se}, which manifests itself as a double copy of Virasoro algebra. The finite dimensional Poincaré algebra is a subalgebra of the extended $\mathfrak{bms}$ algebra. These  infinite dimensional $\mathfrak{bms}$ algebras have gained a renewed importance due to recent developments on the relations between soft theorems and asymptotic symmetries in analyzing the vacuua of gauge theories and gravity \cite{Strominger:2013jfa, He:2014laa, Strominger:2014pwa, Avery:2016zce, Hamada:2018vrw, AtulBhatkar:2019vcb}. It is well understood that at any null boundary in two or three dimensional spacetime one obtain infinite dimensional algebra by constructing the conserved charges without imposing any specific boundary conditions \cite{Adami:2020ugu}. Recently, this is also realized in general dimensions \cite{Adami:2021nnf}.

It is interesting to understand how these $\mathfrak{bms}$ algebras get modified in presence of extended supersymmetries in a theory of gravity. Furthermore in the presence of internal gauge fields, namely the $R-$ symmetry fields, the supercharges non trivially rotate among themselves. This brings interesting dynamics to the system such as the BPS conditions gets modified in presence of the $R-$charges \cite{Ferrara:2010ru, Fuentealba:2017fck}. The effects of extended supersymmetries and $R-$symmetries have been extensively studied in the context of asymptotic symmetries of three dimensional supergravity theories. The supersymmetric deformations of $\mathfrak{bms}_{3}$ algebras and their consequences have been detailed in \cite{Barnich:2014cwa,   Barnich:2015sca, Banerjee:2016nio, Lodato:2016alv, Banerjee:2017gzj, Fuentealba:2017fck, Banerjee:2018hbl, Banerjee:2019lrv, Banerjee:2019epe, Banerjee:2021uxl}. In particular it has been shown that the $R-$charges also get an infinite extensions at the null infinity and the space of the asymptotically flat cosmological solutions gets hugely extended \cite{Banerjee:2018hbl, Fuentealba:2017fck}. A similar study in the context of four dimensional asymptotically flat extended supergravity theory has not been performed yet. This brings us to look for deformations of $\BMSf$ algebra. 

There are two distinct possible ways of generating new algebras starting from one, namely deformation and contraction of an algebra. Deformation of a Lie algebra can be viewed as an inverse procedure of In\"on\"u-Wigner \cite{Inonu:1953sp} contraction.  While physicists have tackled more with contraction of Lie algebras, deformations of various well-known Lie algebras in physics have been recently considered in literature \cite{bacry1968possible, Figueroa-OFarrill:1989wmj, Figueroa-OFarrill:2017ycu,  Figueroa-OFarrill:2018ygf, Enriquez-Rojo:2021rtv, Enriquez-Rojo:2021hna}. In contraction prescription one tries to obtain a new non isomorphic algebra through specific limits of a known algebra, where as in deformation prescription one deforms a Lie algebra to get new (more stable) algebras by turning on structure constants in some commutators \cite{nijenhuis1967deformations,levy1967deformation}. For instance, one may take the limit of the Poincar\'{e} algebra by sending the speed of light to infinity (or to zero) to get Galilean (or Carroll) algebra and conversely the Galilean (or the Carroll) algebra may be deformed into the Poincar\'{e} algebra \cite{levy1965nouvelle,bacry1968possible}. 
In recent works \cite{Parsa:2018kys,Safari:2019zmc}, it has been proven that the three and four dimensional pure $\mathfrak{bms}$ algebras can be deformed, in their non-ideal part, into two families of new non-isomorphic infinite dimensional algebras called $W$ algebras. In the context of three space time dimensions, these are known as $W(a,b)$ algebras, where $\BMSt$ corresponds to $W(0,-1)$ and in the context of four space time dimensions these are knows as $W(a,b;\bar a,\bar b)$ algebras, where $\BMSf$ corresponds to $W(-1/2.-1/2;-1/2,-1/2)$. 
It has been shown that by imposing appropriate boundary conditions $W(0,b)$ as well as $W(b,b;b,b)$ algebras are obtained as near horizon symmetry algebras of 3- and 4-dimensional black-holes \cite{Grumiller:2019fmp}. Also, $W(b,b;b,b)$ has been obtained as asymptotic symmetry algebra of flat Friedmann-Lemaître-Robertson-Walker (FLRW) spacetimes \cite{Enriquez-Rojo:2021blc}. On the other hand, $W(0,1), W(0,0)$ and $W(0,-1)$ have appeared as asymptotic symmetry algebras in various gravitational theories \cite{Afshar:2021qvi, Compere:2013bya, barnich2007classical}. 

In this paper, our goal is to extend the $W(a, b)$ and $W(a, b; \bar{a}, \bar{b})$ algebras with two (fermionic) supercharges. We further consider that the two supercharges rotate among themselves due to an internal R-symmetry.
Our construction is purely group theoretic with only two inputs: a) we demand the consistency of the possible extended algebra with the Jacobi identities and b) we demand that the extended algebra contains Super-Poincar\'{e} algebra as its subalgebra, for particular values of deformation parameters.  The explicit construction goes as follows: 
\begin{itemize}
    \item In three space time dimensions, we first introduce a set of infinite fermionic generators to grade the known $W(a,b)$ algebras and ensure that the resulting superalgebra satisfies graded Jacobi identities. Next we perform the similar construction with infinite bosonic $R-$charge generators. We have further extended our analysis to include the central charges in the algebras. As we have stated above $W(0,-1)$ gives the usual $\BMSt$ and various Supersymmetric extensions of $\BMSt$ algebras are well investigated \cite{Banerjee:2018hbl},\cite{Banerjee:2019lrv}. Our construction of supersymmetric centrally extended $W(a,b)$ algebras in this paper reproduces the known results for $a=0, b=-1$, although we get a new possible central extension.
 
\item 

So far in four space time dimensions we know 
generic bosonic $W(a, b; \bar{a}, \bar{b})$ algebra with central extensions \cite{Safari:2019zmc}. As stated above $\BMSf$ is a special case of these for $a=b=\bar a=\bar b =-1/2.$ Furthermore in \cite{Fotopoulos:2020bqj} minimal supersymmetric generalization of $\BMSf$ with 1 supercharge has been obtained. In this paper, we first extend bosonic $W(a, b; \bar{a}, \bar{b})$ with a set of infinite supercharges. Next we perform further extension with infinite $R-$charges. In this case, the resulting algebra has not been centrally extended. Interestingly we find that for $\BMSf$ with two supercharges, one can not infinitely extend the R charge sectors. We have shown this rigidity for both linear and quadratic dependence of the structure constants. The Jacobi identities only gets satisfied within the global sector, i.e. for $\mathcal{N}=2$ Super-Poincar\'{e} algebra with global $R-$charges. This is one of the prime results of this paper.  
\end{itemize}
Here we must mention that both the three and four dimensional algebras constructed in this paper are purely mathematical. In both cases, the corresponding Super-Poincar\'{e} algebras are embedded in them as subalgebras for appropriate values of the deformation parameters. Thus, in principle, these algebras might show up as the asymptotic symmetry algebras for three and four dimensional asymptotically flat theories. In particular, The ${\cal{N}}=2$ extension of $\BMSf$ is a probable candidate of four dimensional ${\cal{N}}=2$ Supergravity theories with $R-$charges.  

Let us summarize the organisation and results of the paper below :\\
\begin{itemize} 
\item 
In section \ref{section2}, we begin with a brief review of the basic properties of $W(a,b)$ algebra . Next we present new analysis  on a $\mathcal{N}=2$ supersymmetric extension in the presence of $R$ charge which rotate the supercharges among themselves in the rest of this section. We have completed this section with the central extension of the supersymmetric $W(a,b)$ algebra. Equations \eqref{SUSY-W(a,b)} represent the $\mathcal{N}=2, \, W(a,b)$ algebra, where as equations  \eqref{Central-SUSY-W(a,b)} along with the table below them represent its most generic central extensions. 
\item 
Section \ref{sec:bms4-intro} discusses the basic properties of $W(a, b, \bar{a}, \bar{b})$ algebra and may be skipped by the experts. We have added it for establishing the notations used in the later sections. 
\item
In section \ref{sec4}, we extend the $W(a, b, \bar{a}, \bar{b})$ algebra to include two supercharges, albeit any internal symmetry. This section forms the base of the main results of the paper that have been presented in section \ref{sec5}. Equations \eqref{eq:susyW-4d} presents the prime results of this section, which is an infinite extension of $\BMSf$ in presence of two supercharges.
\item 
Section \ref{sec5} contains the most important results of this paper. In this section we have introduced two sets of $R$ charges along with other $\mathcal{N}=2 \,\, W(a, b, \bar{a}, \bar{b})$ generators and studied the possibility to find an infinite extension of the algebra. Here we have considered both the cases for structure constants being linear and non-linear in its arguments and performed a detailed analysis. We find a non-affirmative result (unlike the case of section \ref{sec4}) as discussed in the end of the section. 

\item 
In section \ref{sec6}, we will conclude with a discussion on main results and possible future directions.

\end{itemize}
\section{3-Dimensional Supersymmetric \texorpdfstring{$W(a,b)$}{W(a,b)} and $R$-extended \texorpdfstring{$W(a,b)$}{W(a,b)}}\label{section2}
Earlier works, such as \cite{Parsa:2018kys,Safari:2019zmc} discussed aspects of deformation and stability of $\BMSt$ and $\BMSf$ algebras. In this section, we briefly state their results and observations for $\BMSt$. The centerless $\BMSt$ algebra can be written as
\begin{eqnarray}
\label{WittJ}
\left[J_m,J_n\right]&=&(m-n)J_{m+n}\ ,\\
\left[J_m,P_n\right]&=&(m-n)P_{m+n}\ ,\\
\label{Palg}
\left[P_m,P_n\right]&=&0\ .
\end{eqnarray}
Physically, the $J_m$ s are identified with superrotations while the $P_n$ s are supertranslations. This algebra can be deformed into two parameters family algebra called $W(a,b)$ where $a,b$ are arbitrary real parameters \cite{Safari:2019zmc}. Explicitly, the $W(a,b)$ algebra is given by
\begin{equation} 
\label{eq:W(a,b)-algebra}
\begin{split}
 & [{J}_{m},{J}_{n}]=(m-n){J}_{m+n}\ , \\
 &[{J}_{m},{P}_{n}]=-(n+bm+a){P}_{m+n}\ ,\\
 &[{P}_{m},{P}_{n}]=0.
\end{split}
\end{equation}
It is straightforward to see that $W(0,-1)$ corresponds to $\BMSt$.
\subsection{Supersymmetric $W(a,b)$ algebra}
In this section, we write down a \emph{supersymmetric} version of the $W(a,b)$ algebra. Subsequently, we will introduce $R$ and $S$ charges and also determine the central extension to the algebra. Our approach will be somewhat operational : we start by introducing fermionic generators $G_s$ in the $W(a,b)$ algebra where $s$ runs over half-integers. Our goal would be to write down an extended algebra starting with the centerless $W(a,b)$ algebra as given above by demanding consistency of Jacobi identities. For the time being, unlike $\BMSt$, we do not search for  the realization of the algebra as the asymptotic symmetry algebra of a supersymmetric theory at null infinity in 3 spacetime dimensions. 

Along with the usual $W(a,b)$ algebra as given in \eqref{eq:W(a,b)-algebra}, we introduce the following three commutators
\begin{align}
\label{eq:W(a,b)-extnsm-1}
&\{G_r,G_s\}=P_{r+s}\ ,\\
&[J_m,G_s]=\alpha(m,s)G_{m+s}\ , \label{eq:W(a,b)-extnsm-2}\\
\label{eq:W(a,b)-extnsm-3}
&[P_m, G_s]=\beta(m,s)G_{l(m,s)} \ .
\end{align}
The above extension is motivated by various super-$\BMSt$ algebras written in \cite{Banerjee:2017gzj,Banerjee:2018hbl,Banerjee:2019lrv}. We choose to normalize the super-current generators $G_s$ in a way such that the structure constant appearing in \eqref{eq:W(a,b)-extnsm-1} is unity. It is expected that any deformation of the $\BMSt$ algebra by the parameters $a$ and $b$ will not change the index structure appearing on the RHS of \eqref{eq:W(a,b)-extnsm-2}. For $\BMSt$, it is known that $[P_m,G_r]=0$. However, it is possible that a deformation gives a non-trivial commutator between the supercurrents and supertranslation which vanishes when $a=0, b=-1$\footnote{One can think of the expression $a+b+1$. Clearly for $\BMSt$, this combination identically vanishes and can be a possible candidate structure constant in \eqref{eq:W(a,b)-extnsm-3}. However, if $l(m,s)$ is indeed a linear function of its argument, we will see such a commutator does not satisfy the Jacobi Identity.}. This motivates us to propose \eqref{eq:W(a,b)-extnsm-3} where $l(m,s)$ is a linear function in $m$ and $s$. The structure constants $\alpha$ and $\beta$ appearing above are also assumed to be linear functions of its arguments. Our strategy will be to fix these structure constants and $l(m,s)$ by demanding consistency of certain relevant Jacobi identities. 

The Jacobi identity involving the generators $G_r, G_s$ and $P_m$ is given by
\begin{equation}
    [\{G_s,G_r\},P_m]=\{G_s,[G_r,P_m]\}+\{G_r,[G_s,P_m]\}\ .
\end{equation}
Using \eqref{eq:W(a,b)-algebra} and \eqref{eq:W(a,b)-extnsm-1}-\eqref{eq:W(a,b)-extnsm-3}, we get,
\begin{equation}
    \beta(m,r)P_{s+l(m,r)}+\beta(m,s)P_{r+l(m,s)}=0
\end{equation}
Assuming linearity of $l(m,s)$ and the structure constant $\beta(m,s)$ in both of their arguments,  the above equation is satisfied if $l(m,s)=l_0+l_1m +s$ where $l_0$ and $l_1$ are constants and $\beta(m,r)=-\beta(m,s)$ for any $r,s$. This hence yields $\beta(m,s)=0$.

Next, we use Jacobi identities on the operators $J_m, G_s$ and $G_r$ to determine $\alpha(m,s)$. The corresponding Jacobi identity is
\begin{equation}
    [ \{G_s,G_r\},J_m ] =\{G_s,[G_r,J_m]\}+\{G_r,[G_s,J_m]\}\ .
\end{equation}
Using \eqref{eq:W(a,b)-algebra}, \eqref{eq:W(a,b)-extnsm-1}-\eqref{eq:W(a,b)-extnsm-3} we get
\begin{equation}
-\alpha(m,r)P_{m+r+s}-\alpha(m,s)P_{m+r+s}=(r+s+bm+a)P_{m+r+s}\ .
\end{equation}
Clearly, equating the coefficients of $P_{m+r+s}$, we recover,
\begin{equation}
\label{eq:J-G-structureconstant}
    \alpha(m,s)=-\left(\frac{bm+a}{2}+s\right)\ .
\end{equation}
The Jacobi identity applied on $J_m,J_n$ and $G_s$ reads as
\begin{equation}
   [[G_s,J_m],J_n]+[[J_m,J_n],G_s]+[[J_n,G_s],J_m]=0\ .
\end{equation}
A similar exercise on the above Jacobi identity followed by equating the coefficient of $G_{m+n+s}$ yields
\begin{equation}
\alpha(m,s)\alpha(n,m+s)+(m-n)\alpha(m+s,s)=\alpha(n,s)\alpha(m,n+s)\ .
\end{equation}
It can be easily seen that the structure constant $\alpha(m,s)$ as determined in \eqref{eq:J-G-structureconstant} indeed satisfies the above equality. We thus end up with a possible $\mathcal{N}=1$ supersymmetric extension of $W(a,b)$ algebra given by
\begin{align}
  & [{J}_{m},{J}_{n}]=(m-n){J}_{m+n}, \\
 &[{J}_{m},{P}_{n}]=-(n+bm+a){P}_{m+n},\\
 &[{P}_{m},{P}_{n}]=0,\\
 &\{G_r,G_s\}=P_{r+s},\\
 &[J_m,G_s]=-\left(\frac{bm+a}{2}+s\right)G_{m+s},\\
 &[P_m, G_s]=0
\end{align}

Now that we have obtained a possible $\mathcal{N}=1$ extension of the $W(a,b)$ algebra, we can consider including another copy of fermionic supercharges which we denote by $H_s$ where the index $s$ can take half-integer values. They satisfy the following commutators with the superrotation and supertranslation generators of the $W(a,b)$ algebra 
\begin{align}
  &\{H_r,H_s\}=P_{r+s},\quad [P_m, H_s]=0\ ,\\
  &[J_m,H_s]=-\left(\frac{bm+a}{2}+s\right)H_{m+s}\ .
\end{align}
The supercurrent generators $G_s$ and $H_s$ can be used to define the following linear combinations
\begin{align}
    Q_r^1=\frac{1}{2}\left(G_r+iH_r\right)\ ,\quad Q_r^2=\frac{1}{2}\left(G_r-iH_r\right)\ .
\end{align}
The newly defined generators $Q^1_r$ and $Q^2_r$ satisfy
\begin{align}
\label{eq:G-H-lin-combination}
    &\{Q_r^1,Q_s^2\}=P_{r+s}\ ,\quad \{Q_r^1,Q_s^1\}=0\ , \quad \{Q_r^2,Q_s^2\}=0\ ,\notag\\
    &[J_m,Q^i_s]=-\left(\frac{bm+a}{2}+s\right)Q^i_{m+s}\ , \quad [P_m,Q^i_s]=0\ .
\end{align}
\subsection{R-extension of Supersymmetric $W(a,b)$ algebra}
Our next aim is to write the generalized algebra in the presence of $R$-charges. Physically speaking, the $R$-charge generators rotate the supercharge generators and thus brings in an additional non trivialities in the algebra. It is known that introduction of $R$ charge generators necessitates the introduction of $S$-charge generators \cite{Howe:1995zm} in the context of $\BMSt$. Motivated by the $\mathcal{N}=2$ $\BMSt$ algebra as discussed in \cite{Banerjee:2019lrv} we begin our analysis by proposing the following relations involving the $R$-charge and $S$-charge generators
\begin{eqnarray}
\label{eq:R-Q-commutator}
    [R_n, Q_r^1]=\beta(n,r)~Q_{n+r}^1,&\quad & [R_n, Q_r^2]=-\beta'(n,r)Q_{n+r}^2\ ,\\
\label{eq:Q-Q-anticommutator}
    {[P_n,R_m]}=\sigma(n,m)S_{n+m}, &\quad & \{Q_r^1,Q_s^2\}=P_{r+s}+\eta(r,s)S_{r+s}\ ,\\
\label{eq:RJ-SJ-commutator}
    {[R_n,J_m]}=\gamma(n,m)R_{n+m}, &\quad & [S_n,J_m]=\kappa(n,m)S_{n+m}\ .
\end{eqnarray}

In writing the above ansatz, we have assumed that an $R-$deformation of the $\mathcal{N}=2$ super-$W(a,b)$ algebra will not affect the index structure of the undeformed algebra. The Jacobi identity for the operators $Q_m^1, Q_n^2$ and $R_s$ is given by,
\begin{equation}
\label{eq:QQR-Jacobi}
[ \{Q_r^1, Q_s^2\},R_m ] =\{Q_r^1,[Q_s^2,R_m]\}+\{Q_s^2,[Q_r^1,R_m]\}\ .
\end{equation}
Using \eqref{eq:R-Q-commutator} and \eqref{eq:Q-Q-anticommutator} in the above, we get,
\begin{equation}
\label{eq:QQR-Jacobi-result}
\begin{aligned}
    {[P_{r+s}+\eta(r,s)S_{r+s},R_m]}&=(\beta'(m,s)-\beta(m,r))P_{r+s+m}\\
    &\hspace{0.5cm}+(\beta'(m,s)\eta(r,m+s)-\beta(m,r)\eta(m+r,s))S_{m+r+s}\ .
\end{aligned}
\end{equation}

Further, noting that $[S_m,R_n]= 0$ implies that the LHS of \eqref{eq:QQR-Jacobi-result} is independent of the translation generator $P_m$. In order to make this Jacobi identity consistent the coefficient of the translation generator on RHS must vanish identically implying
\begin{equation}
    \beta'(m,s)=\beta(m,r)\ .
\end{equation}
Since the above has to be true for arbitrary half-integer values of $r,s$ and integer values of $m$ and also both $\beta$ and $\beta'$ must be linear, we conclude that both $\beta'(m,s)$ and $\beta(m,r)$ depend only on $m$. For simplicity, we denote the structure constant appearing in \eqref{eq:R-Q-commutator} as $\beta(m)$ since,
\begin{equation}
    \beta'(m)=\beta(m)\ .
\end{equation}

Demanding further consistency of  \eqref{eq:QQR-Jacobi-result} yields,
\begin{align}
\label{eq:sigma-eta-relation}
    \sigma(r+s,m)= \beta(m)(\eta(r,s+m)-\eta(r+m,s))\ .
\end{align}
The two Jacobi identities involving $Q_r^1, J_n$ and $R_m$ and,  $R_l, J_m$ and $J_n$,
\begin{eqnarray}
[[Q_r^1, J_n], R_m] + [[J_n, R_m], Q_r^1] + [[R_m, Q_r^1], J_n]&=&0\ ,\\
{[[R_l, J_m], J_n]}+[[J_m, J_n], R_l]+[[J_n, R_l],J_m]&=&0\ ,
\end{eqnarray}
 lead to
\begin{eqnarray}
\label{eq:beta-gamma-relation}
    &m\beta(m)=\gamma(m,n)\beta(m+n)\ ,&\\
  \label{eq:RJJ}
    &\gamma(l,m)\gamma(l+m,n)-(m-n)\gamma(l, m+n)-\gamma(l,n)\gamma(l+n, m)&=0\ . 
\end{eqnarray}
Assuming the structure constants $\gamma(m,n)$ and $\beta(m)$ to be linear in $m$ and $n$ we explicitly take them to be of the form,
\begin{eqnarray}
\label{eq:betaansatz}
\beta(m)&=&\beta_0 +m\beta_1\ ,\\
\label{eq:gammaansatz}
\gamma(m,n)&=&\gamma_0+\gamma_1 m+\gamma_2 n\label{gam}\ .
\end{eqnarray}
The above ansatz for $\gamma(m,n)$ along with \eqref{eq:RJJ} ensures
\begin{align}
    \gamma_1=1\ .
\end{align}
\eqref{eq:beta-gamma-relation} then gives us five equations satisfied by four parameters $\beta_0,\beta_1, \gamma_0$ and $\gamma_2$ are
\begin{eqnarray}
\gamma_0\beta_0 = 0\ , & \quad & \beta_0 \gamma_2+\beta_1 \gamma_0=0\ ,\\
\gamma_0 \beta_1=0\ , & \quad & \beta_1(\gamma_2+1)=0\ ,\\
\gamma_2 \beta_1 &=&0\ .
\end{eqnarray}
Clearly the above set of equations are over constrained but admits the following consistent solution
\begin{equation}
    \gamma_0 =0\ ,\quad \gamma_2=0\ ,\quad \beta_1=0\ ,
\end{equation}
while $\beta_0$ is a non-zero constant that cannot be fixed further. One can easily check that this is also consistent with the Jacobi identity for $R_m, R_n$ and $Q^i_r$. Thus, we can write the following commutation relations 
\begin{equation}\label{RQJ}
    [R_n,Q^1_r]=\beta_0 Q^1_{n+r}\ , \quad [R_n,Q^2_r]=-\beta_0 Q^2_{n+r}\ , \quad [R_n,J_m]=nR_{n+m}\ .
\end{equation}
The Jacobi identity involving $J_m, J_n$ and $S_l$, 
\begin{equation}
    [[J_m, J_n], S_l]+[[J_n, S_l],J_m]+[[S_l, J_m],J_n]=0
\end{equation}  
yields
\begin{equation}
\label{eq:JJS}
\kappa(l,m)\kappa(l+m,n)-(m-n)\kappa(l,m+n)-\kappa(l,n)\kappa(n+l,m)=0\ .
\end{equation}
Similar to the ansatz for $\gamma(m,n)$, we assume the following anstaz for $\kappa(m,n)$
\begin{equation}
\label{eq:kappa}
    \kappa(m,n)=\kappa_0+\kappa_1 m+\kappa_2 n .
\end{equation}
Plugging  the above ansatz into \eqref{eq:JJS} we get
\begin{equation}
    \kappa_1=1\ .
\end{equation}
The Jacobi identity for the operators $J_l, P_m$ and $R_n$,
\begin{align}
    [[J_l , P_m] , R_n]+[[P_m,R_n],J_l]+[[R_n, J_l], P_m]=0\ ,
\end{align}
leads to the relation
\begin{align}
\label{eq:sigma-kappa-relation}
    -(m+b l+a) \sigma(m+l,n)+\sigma(m,n)\kappa(m+n,l)-n\sigma(m,n+l)=0\ .
\end{align}
Assuming a linear ansatz for $\sigma(m,n)$ i.e.
\begin{equation}
    \sigma(m,n)=\sigma_0+\sigma_1 m+\sigma_2 n\ ,
\end{equation}
one can substitute it back into \eqref{eq:sigma-kappa-relation} to get the set of seven relations:
\begin{eqnarray}
\label{eq:sigma-kappa-system1}
b\sigma_1 =0\ ,&\quad &\sigma_0(\kappa_0-a)=0\ ,\\
\label{eq:sigma-kappa-system2}
\sigma_1(\kappa_2-b-1)=0\ ,&\quad&\sigma_2(\kappa_2-b-1)=0\ ,\\
\label{eq:sigma-kappa-system3}
\sigma_1(\kappa_0-a)=0\ ,&\quad&\sigma_2(\kappa_0-a)=0\ ,\\
\label{eq:sigma-kappa-system4}
\sigma_0(\kappa_2-b)-a\sigma_1&=&0\ .
\end{eqnarray}
The above system of equations have \textbf{two} consistent solution sets (detailed in appendix \ref{app:sigma-kappasoln}):
\begin{itemize}\label{cases}
    \item Case I: $\sigma_0=\sigma_1=0$; $\kappa_0=a$, $\kappa_2=b+1$ while $\sigma_2$ is arbitrary.
    \item Case II: $\sigma_1=\sigma_2=0$; $\kappa_0=a$, $\kappa_2=b$ while $\sigma_0$ is arbitrary.
\end{itemize}
In light of the above, we can rewrite \eqref{eq:Q-Q-anticommutator} and \eqref{eq:RJ-SJ-commutator} as
\begin{equation}
\begin{aligned}
    \text{Case I:}& \quad [P_n,R_m]=\sigma_2 mS_{n+m}\ , \quad [S_n,J_m]=(a+n+(b+1)m)S_{n+m}\ ;\\
    \text{Case II:}& \quad [P_n,R_m]=\sigma_0 S_{n+m}\ , \quad [S_n,J_m]=(a+n+bm)S_{n+m}\ .
\end{aligned}
\end{equation}
Finally, we need to find the structure constant $\eta(r,s)$ appearing in $\{Q_r^1,Q_s^2\}$ in \eqref{eq:Q-Q-anticommutator}.
Assuming a linear form of $\eta(r,s)$ i.e. 
\begin{equation}
\label{eq:eta-ansatz}
    \eta(r,s)=\eta_0+\eta_1 r +\eta_2 s\ ,
\end{equation}
we use \eqref{eq:sigma-eta-relation} to see that we must have a relation of the form
\begin{equation}
    \sigma(r+s,m)=m\beta_0(\eta_2-\eta_1)\ .
\end{equation}
It is quite evident that a choice of parameters as defined in Case II in our preceding analysis is inconsistent with the above equation since the LHS is a constant and independent of $m$. However, from the structure constants of Case I, we arrive at the relation
\begin{equation}
    \sigma_2=\beta_0(\eta_2-\eta_1)\ .
\end{equation}
Finally, the Jacobi identity for $Q_r^1, Q_s^2$ and $J_m$ can be written as
\begin{equation}
    [\{Q_r^1,Q_s^2\},J_m]=\{Q_r^1,[Q_s^2,J_m]\}+\{Q^2_s,[Q_r^1,J_m]\}\ .
\end{equation}
Using \eqref{eq:Q-Q-anticommutator} and \eqref{eq:RJ-SJ-commutator}, the above gives rise to,
\begin{equation}
    \eta(r,s)\kappa(r+s,m)=\left(\frac{bm+a}{2}+s\right)\eta(r,s+m) +\left(\frac{bm+a}{2}+r\right)\eta(r+m,s)\ .
\end{equation}
Assuming, $\kappa(m,n)=a+m+(b+1)n$, the above equation will be satisfied for an ansatz of the form \eqref{eq:eta-ansatz} provided,
\begin{equation}
    \eta_0=0 \quad \text{and} \quad \eta_1=-\eta_2\ .
\end{equation}
Thus, the $R$ and $S$-charge sector algebra under a deformation reads as \footnote{$\eta_1=0$ or $\beta_0=0$ are also viable choices of parameters that satisfy the Jacobi identities. But for interpreting $R_0$ as the $R-$symmetry generator we must consider non-zero values of those parameters.}:
\begin{eqnarray}
    [R_n, Q_r^1]=\beta_0 Q_{n+r}^1,&\quad & [R_n, Q_r^2]=-\beta_0 Q_{n+r}^2\ ,\\
    {[P_n,R_m]}=-2\beta_0 \eta_1 mS_{n+m}, &\quad & \{Q_r^1,Q_s^2\}=P_{r+s}+\eta_1 (r- s)S_{r+s}\ ,\\
    {[R_n,J_m]}=nR_{n+m}, &\quad & [S_n,J_m]=(a+n+(b+1)m)S_{n+m}\ .
\end{eqnarray}
Redefining $S_n \rightarrow \mathcal{S}_n/\eta_1$ and $R_n \rightarrow \beta_0 \mathcal{R}_n$, we arrive at,
\begin{eqnarray}
    [\mathcal{R}_n, Q_r^1]=Q_{n+r}^1,&\quad & [\mathcal{R}_n, Q_r^2]=- Q_{n+r}^2\ ,\\
    {[P_n,\mathcal{R}_m]}=-2 m \mathcal{S}_{n+m}, &\quad & \{Q_r^1,Q_s^2\}=P_{r+s}+ (r-s)\mathcal{S}_{r+s}\ ,\\
    {[\mathcal{R}_n,J_m]}=n\mathcal{R}_{n+m}, &\quad & [\mathcal{S}_n,J_m]=(a+n+(b+1)m)\mathcal{S}_{n+m}\ .
\end{eqnarray}
The full $W(a,b)$ algebra including $R$ and $S$ charge generators takes the form
\begin{align}\label{SUSY-W(a,b)}
  & [{J}_{m},{J}_{n}]=(m-n){J}_{m+n},\quad [{J}_{m},{P}_{n}]=-(n+bm+a){P}_{m+n},\notag\\ 
  &[J_m,Q^1_r]=-\left(\frac{bm+a}{2}+r\right)Q^1_{m+r}, \quad [J_m,Q^2_r]=-\left(\frac{bm+a}{2}+r\right)Q^2_{m+r},
  \notag\\
    &[J_m,\mathcal{R}_n]=-n\mathcal{R}_{n+m},\notag\\  &[J_m,\mathcal{S}_n]=-(a+n+(b+1)m)\mathcal{S}_{n+m},\notag\\
  & [\mathcal{R}_m, Q_r^1]= Q_{m+r}^1,\quad  [\mathcal{R}_m, Q_r^2]=- Q_{m+r}^2, \notag\\
  & \{Q_r^1,Q_s^2\}=P_{r+s}+(r-s)\mathcal{S}_{r+s},\quad [P_m,\mathcal{R}_n]=-2nS_{n+m}.
\end{align}
where indices $\{m,n,p,q\} \in \mathbb{Z}$ while $\{r,s\} \in \mathbb{Z}\half$ and $i\in \{1,2\}$. All other commutators vanish. The conformal weight of the generators $P_m$ and $S_m$ are $-b+1$ and $-b$ respectively, while the weight of $Q^1,Q^2$ is $-\frac{b}{2}+1$. For the specific case $a=0$ and $b=-1$ which corresponds to supersymmetric-$\BMSt$, we recover the same algebra as given in \cite{Banerjee:2019lrv} {\footnote{There is a typo in Eq. 3.19 and Eq. 8.59 of \cite{Banerjee:2019lrv}. The structure constant in $[M_n, R_m]$ commutator will be $-2m$ instead of $-4m$ otherwise the $({\cal G}_r^1, {\cal G}_s^2, {R}_{m})$ will not be satisfied.}}.

\subsection{Central extensions of supersymmetric \texorpdfstring{$W(a,b)$}{W(a,b)}}

One can show that the $W(a,b)$ algebra for generic values of its parameters just admits one central term in its Witt part but for certain specific values of $a$ and $b$, it admits various central extensions which were classified in \cite{gao2011low}. The most general centrally extended supersymmetric $W(a,b)$ algebra can be written as, 
\begin{align}
\label{eq:centralextansatz1}
  & [{J}_{m},{J}_{n}]=(m-n){J}_{m+n}+u(m,n),\\ &[{J}_{m},{P}_{n}]=-(n+bm+a){P}_{m+n}+v(m,n),\\ 
  &[J_m,Q^1_r]=-\left(\frac{bm+a}{2}+r\right)Q^1_{m+r}+x_1(m,r),\\
  \label{eq:centraltermJQ2}
  &[J_m,Q^2_r]=-\left(\frac{bm+a}{2}+r\right)Q^2_{m+r}+x_2(m,r),\\
    &[J_m,\mathcal{R}_n]=-n\mathcal{R}_{n+m}+y(m,n),\\& [J_m,\mathcal{S}_n]=-(a+n+(b+1)m)\mathcal{S}_{n+m}+z(m,n),\\
  & [\mathcal{R}_m, Q_r^1]= Q_{m+r}^1+g_{1}(m,r),\quad  [\mathcal{R}_m, Q_r^2]=- Q_{m+r}^2+g_{2}(m,r), \\
  & \{Q_r^1,Q_s^2\}=P_{r+s}+(r-s)\mathcal{S}_{r+s}+f(r,s),\quad [P_m,\mathcal{R}_n]=-2n\mathcal{S}_{n+m}+h(m,n),\\
  &\{Q_r^1,Q_s^1\}=w_{1}(r,s), \quad \{Q_r^2,Q_s^2\}=w_{2}(r,s), \\
  & [{P}_{m},{P}_{n}]=t_1(m,n),\ [\mathcal{R}_{m},\mathcal{R}_{n}]=w(m,n),\\& [\mathcal{R}_m,\mathcal{S}_n]=k(m,n),\ [\mathcal{S}_m,\mathcal{S}_n]=s(m,n)\\
  \label{eq:centralextansatz9}
  &[P_m, \mathcal{S}_{n}]=t_2(m,n),\ [P_m,Q_r^i]=h^{i}(m,r),\ [\mathcal{S}_{n},Q_r^i]=f^{i}(n,r)\ ,
\end{align}
where the Jacobi identity between the generators are expected put constraints on unknown functions $u,v,x_i,y,z,w,g_i,f,h,w_i,t_i,k,s,h^i,f^i$, that denote the possible central extensions.

The central term in the commutator $[{J}_{m},{J}_{n}]$ which we denoted as $u(m,n)$, is an arbitrary anti symmetric function. The Jacobi identity 
\begin{equation}
[J_m,[J_n,J_l]]+[J_n,[J_l,J_{m}]]+[J_l,[J_m,J_n]]=0,
\end{equation}
leads to the relation, 
\begin{equation}
    (n-l)u(m,n+l)+(l-m)u(n,m+l)+(m-n)u(l,n+m)=0\ ,
\end{equation}
which has the nontrivial solution $u(m,n)=C_{jj}^{(1)}\,(m^3-m)\,\delta_{m+n,0}$. This, as expected, is of the form of the usual Virasoro central charge. Other Jacobi identities do not put any new constraint on $u(m,n)$. A redefinition of $J_m \rightarrow J_m +A \delta_{m,0}$ with an appropriate choice of $A$ can be used to absorb the linear term in $m$.

One can fix the central term $v(m,n)$ of the $[{J}_{m},{P}_{n}]$ commutator in the following way. The Jacobi identity between $J_m, J_n$ and $P_l$ leads to
\begin{equation}\label{v-central}
    -(a+bn+l)v(m,n+l)+(a+bm+l)v(n,m+l)-(m-n)v(n+m,l)=0\ .
\end{equation}
Specific values of $a$ and $b$, yield even more non-trivial solutions. We systematically tabulate all the cases below
\begin{enumerate}
    \item $a=b=0$ where, $v(m,n)=(C^{(1)}_{jp} m^2 + C^{(2)}_{jp} m)\delta_{m+n,0}$\ ,
    \item $a=0$, $b=1$ where, $v(m,n)= (C^{(3)}_{jp}m+C^{(4)}_{jp})\delta_{m+n,0}$\ ,
    \item $a=0$, $b=-1$ where, $v(m,n)= (C^{(5)}_{jp}m^3+C^{(6)}_{jp}m)\delta_{m+n,0}$\ ,
    \item $a=0$, $b \neq 0, 1,-1$ where, $v(m,n)=C_{jp}^{(7)}m\delta_{m+n,0}$\ ,
    \item $a \neq 0$ and $b$ is arbitrary where, $v(m,n)=C_{jp}^{(8)}\left(1+\frac{b-1}{a}m\right)\delta_{m+n,0}$\ .
\end{enumerate}
Here the subscript $jp$ denotes the central extension in $[J,P]$ commutator. Out of the 8 central terms appearing in the above five scenarios only $C^{(1)}_{jp},C^{(3)}_{jp}, C^{(4)}_{jp}$ and $C^{(5)}_{jp}$ are the non trivial ones. Other central terms can be absorbed by a simple redefinition of ${P}_{m}\rightarrow {P}_{m}+B\delta_{m,0} $ and choosing the constant $B$ subsequently in an appropriate manner. Thus we drop the remaining central terms $C^{(2)}_{jp},C^{(6)}_{jp}, C^{(7)}_{jp}$ and $C^{(8)}_{jp}$ for the remaining analysis.

The above analysis demonstrates that there may be certain values for the parameters $a$ and $b$ for which certain central terms will be allowed in the algebra. This opens up  a host of possibilities in the central extension. We will focus on the most general extension that is admissible for arbitrary values of $a$ and $b$.

The commutator $[J_m,Q^1_r]$ may admit a central term given by, 
\begin{equation}
    [{J}_{m},Q^1_r]=-\left(\frac{bm+a}{2}+r\right)Q^1_{m+r}+x_1(m,r),
\end{equation}
where $x_1(m,r)$ is an arbitrary function. The Jacobi identity between $J_m, J_n$ and $Q^1_r$ gives us,
\begin{equation}
    -\left(\frac{bn+a}{2}+r\right)x_1(m,n+r)+\left(\frac{bm+a}{2}+r\right)x_1(n,m+r)-(m-n)x_1(n+m,r)=0\ .
\end{equation}
The $x_1$ central term appearing in the $[J_m,,Q^1_r]$ commutator is identically zero since a central term proportional to $\delta_{m+r,0}$ is identically zero as $m$ is an integer and $r$ is a half-integer. An identically similar argument is true for $x_2(m,r)$ which is the central extension in the $[J_m,Q^2_r]$ commutator as given in \eqref{eq:centraltermJQ2}. 

The central term in $[J_m,S_n]$ commutator is denoted by $z(m,n)$ and the full commutator is written as 
\begin{equation}
\label{eq:JScentral}
   [J_m,\mathcal{S}_n]=-(a+n+(b+1)m)\mathcal{S}_{n+m}+z(m,n)\ .
\end{equation}
The Jacobi identity of $J_m, J_n$ and $\mathcal{S}_l$ yields,
\begin{equation}
\label{eq:Centralzeqn}
    -(a+(b+1)n+l)z(m,n+l)+(a+(b+1)m+l)z(n,m+l)-(m-n)z(n+m,l)=0,
\end{equation}
which admits the following nontrivial solutions
\begin{enumerate}
    \item $a \neq 0$ and $b$ is arbitrary where $z(m,n)=C^{(0)}_{js}\left(1+\frac{b}{a}m \right)\delta_{m+n,0}$\ ,
    \item $a=0$ and $b \neq 0,-1,-2$ where $z(m,n)=C_{js}^{(1)}m\delta_{m+n,0}$\ ,
    \item $a=0$ and $b=-1$ where $z(m,n)=(C_{js}^{(2)}m+C_{js}^{(3)}m^2)\delta_{m+n,0}$\ ,
    \item $a=0$ and $b=-2$ where $z(m,n)=(C_{js}^{(4)}m+C_{js}^{(5)}m^3)\delta_{m+n,0}$\ ,
    \item $a=b=0$, where $z(m,n)=(C^{(6)}_{js}+C^{(7)}_{js}m)\delta_{m+n,0}$\ .
\end{enumerate} 
Again, performing the shift $\mathcal{S}_m \rightarrow \mathcal{S}_m + S\delta_{m,0}$ will remove some of the constants appearing above with an appropriate choice of $S$. A detailed analysis reveals that we can drop $C_{js}^{(0)}, C_{js}^{(1)}, C_{js}^{(2)}$ and $C_{js}^{(4)}$.

The central term $y(m,n)$, 
\begin{equation}
   [J_m,\mathcal{R}_n]=-n\mathcal{R}_{n+m}+y(m,n),
\end{equation}
can be determined from the Jacobi identity of $J_m, J_n$ and $\mathcal{R}_l$ which gives, 
\begin{equation}
    -ly(m,n+l)+ly(n,m+l)-(m-n)y(n+m,l)=0\ .
\end{equation}
This has the nontrivial solution $y(m,n)=C^{(0)}_{jr} m^2\delta_{m+n,0}$. This was discussed in earlier works \cite{gao2011low, Safari:2019zmc}. 

The anticommutator $ \{Q_r^1,Q_s^2\}$ may have the possible central term $f(r,s)$ and is given by,
\begin{align}
     \{Q_r^1,Q_s^2\}=P_{r+s}+(r-s)\mathcal{S}_{r+s}+f(r,s)\ .
\end{align}
The Jacobi identity of $Q^1_r, Q^2_s$ and $J_m$ gives,
\begin{equation}
    \left(\frac{bm+a}{2}+s\right)f(r,s+m)+\left(\frac{bm+a}{2}+r\right)f(r+m,s)=-v(m,r+s)-(r-s)z(m,r+s)\ .
\end{equation}
The above equation needs to be dealt with care in case-by-case basis. We tabulate all possible solutions for various values of the deformation parameter $a$ and $b$
\begin{enumerate}
    \item When $a=b=0$, we get the solution as $f(r,s)=C_{qq}^{(0)}r \delta_{r+s,0}$. Also, for consistency of the above equation, we must have $C_{jp}^{(1)}=C_{js}^{(6)}=C_{js}^{(7)}=0$. This in turn ensures that for $a=b=0$, $v(m,n)=z(m,n)=0$. Note that  the linear term appearing  in $f(r,s)$ cannot be absorbed in the shift of generators. A possible absorbing of the central term can be performed by shifting the supertranslation generators $P_n$. This was already performed earlier to ensure $C_{jp}^{(2)}$ drops out in the expression of the central charge. Thus, there is no more freedom to absorb this piece in the generators.
    \item When $a=0$ and $b=1$, we get the solution $f(r,s)=C_{qq}^{(1)}\delta_{r+s,0}$. Again, consistency demands us to set $C_{jp}^{(3)}=C_{jp}^{(4)}=0$, again ensuring that $v(m,n)$ vanishes for this case.
    \item When $a=0$ and $b=-1$, we get the solution $f(r,s)=2C_{jp}^{(5)}r^2 \delta_{r+s,0}$ along with the constraint that $C_{js}^{(3)}=0$. This implies for $a=0$ and $b=-1$, we have $z(m,n)=0$.
    \item When $a=0$ and $b=-2$, we get the solution to be $f(r,s)=C_{js}^{(5)}r^3 \delta_{r+s,0}$.
    \item When $a=0$ and $b =2$, we get the solution $f(r,s)=C_{qq}^{(2)}r^2\delta_{r+s,0}$ while for $a=0$ and $b \neq -2,-1,0,1,2$, $f(r,s)$ must vanish identically.
    \item When $a \neq 0$ and $b$ is arbitrary, we recover we get $f(r,s)=0$ identically.
\end{enumerate}

$g_1(m,n)$ is an arbitrary symmetric function which denotes the central term in the $[\mathcal{R}_m, Q_r^1]$ commutator and is given by,
\begin{align}
    [\mathcal{R}_m, Q_r^1]= Q_{m+r}^1+g_{1}(m,r)\ .
\end{align}
For $g_1(m, r) \propto \delta_{m+r,0}$, we can easily conclude that this will be zero identically since $m$ is an integer and $r$ is a half-integer. 

The commutator of $[P_m,\mathcal{R}_n]$ may admit a central term $h(m,n)$ which appears as follows, 
\begin{align}
    [P_m,\mathcal{R}_n]=-2n\mathcal{S}_{m+n}+h(m,n)\ .
\end{align}
The Jacobi identity of $J_m, \mathcal{R}_n$ and $P_l$ leads to
\begin{equation}
\label{eq:hstrconstant}
    (bm+a+l)h(l+m,n)+nh(l,m+n)= 2nz(m,n+l)\ .
\end{equation}
Clearly, the RHS of the above equation depends crucially on the values of $a$ and $b$. However, dealing case-by-case it turns out that $h(m,n)=0$ for all values of $a$ and $b$ along with the constraint $C_{js}^{(5)}=0$. This in turn implies $z(m,n)$ vanishes for $a=0$ and $b=-2$.

The commutator of $[\mathcal{R}_m,\mathcal{S}_n]$ can admit a central extension given by 
\begin{equation}
    [\mathcal{R}_m,\mathcal{S}_n]=k(m,n)\ .
\end{equation}
The Jacobi identity of $Q^1_r, Q^2_s$ and $\mathcal{R}_m$ gives
\begin{equation}
\label{eq:kstrconstant}
    (r-s)k(m,r+s)+f(r,m+s)-f(m+r,s)=0\ .
\end{equation}
Depending on the form of $f(r,s)$, we will have different solutions for $k(m,n)$. We list the possible solutions as follows:
\begin{enumerate}
    \item For $a=b=0$, the above equation simplifies to 
    \begin{equation}
        (r-s)k(m,r+s)=C_{qq}^{(0)}m \delta_{m+r+s,0}\ .
    \end{equation}
    For the above equation to be consistent we must have $C_{qq}^{(0)}=0$ which further implies for this case, $f(r,s)=k(m,n)=0$.
    \item For $a=0, b=1$ we simply recover $(r-s)k(m,r+s)=0$ which immediately implies $k(m,n)=0$.
    \item For $a=0$, $b=-1$, we see that the equation for $k(m,n)$ is satisfied provided $k(m,n)=2C_{jp}^{(5)}m \delta_{m+n,0}$. 
    \item For $a=0$, $b=-2$, consistency demands us to set $C_{js}^{(5)}=0$ which inturn ensures $z(m,n)=f(r,s)=k(m,n)=0$.
    \item For $a=0$, $b=2$, we see the solution of $k(m,n)=C_{qq}^{(2)}m \delta_{r+s,0}$.
    \item For $a \neq 0$ and arbitrary $b$, we must have $k(r,s)=0$.
\end{enumerate}

A quick glance at \eqref{eq:centralextansatz1}-\eqref{eq:centralextansatz9} tells us that the central extension to the commutator $[\mathcal{R}_m,\mathcal{R}_n]$ will affect the Jacobi identity between $J_m$, $\mathcal{R}_n$ and $\mathcal{R}_p$. Denoting the central extension in this case as
\begin{equation}
    [\mathcal{R}_m, \mathcal{R}_n]=w(m,n)\ ,
\end{equation}
the $J_m$, $\mathcal{R}_n$, $\mathcal{R}_p$ Jacobi identity leads to
\begin{equation}
    pw(n,p+m)=nw(p,m+n)\ .
\end{equation}
A little algebra shows that the solution to the above functional equation is given by $w(m,n)=C_{rr}m\delta_{m+n,0}$.

The central term in the $[S_m,S_n]$ commutator is denoted as $s(m,n)$ and can be explicitly written as  
\begin{equation}
   [S_m,\mathcal{S}_n]=s(m,n)\ .
\end{equation}
The Jacobi identity between $P_m, \mathcal{R}_n$ and $\mathcal{S}_l$ gives
\begin{equation}
    ns(l,m+n)=0\ ,
\end{equation}
which naturally implies $s(m,n)=0$ identically. The reader can easily verify that the Jacobi identities of ($\mathcal{R}_m, \mathcal{S}_n, Q^i_r$), ($\mathcal{R}_m, P_n, Q^i_r$) and ($Q^1_r, Q^2_s, P_m $) implies $f^i(m,n)=h^i(m,n)=t_2(m,n)=0$. The supertranslation commutator $[P_m,P_n]$ also does not admit any central term, as discussed in further details in an earlier work \cite{Parsa:2018kys} by one of the authors. 

Finally, we consider the anticommutator $\{Q_r^1,Q_s^1\}$ which may admit a central term as 
\begin{align}
     \{Q_r^1,Q_s^1\}=w_{1}(r,s)\ .
\end{align}
It is clear that $w_1(r,s)$ must be symmetric in its arguments. The Jacobi identity between $Q^1_r, Q^1_s$ and $R_m$ gives
\begin{equation}
    w_1(r,s+m)+w_1(s,m+r)=0\ .
\end{equation}
For $m=0$, we see that $w_1(m,n)$ should be antisymmetric which is clearly a contradiction. Thus, $w_1(m,n)=0$ identically and a similar argument involving $Q^2_r$ yields $w_2(m,n)=0$. This completes a full description of the central extension for the $W(a,b)$ algebra which clearly depends on the values of the parameters $a$ and $b$. 

Here, we tabulate the $W(a,b)$ for $a= 0$ and $b=-1$, which is the same as the super-$\BMSt$ algebra 
\begin{equation}
\begin{aligned}
\label{Central-SUSY-W(a,b)}
  & [{J}_{m},{J}_{n}]=(m-n){J}_{m+n}+C_{jj}^{(1)}\,m^3\,\delta_{m+n,0},\\ & [{J}_{m},{P}_{n}]=-(n-m){P}_{m+n}+C_{jp}^{(5)}\,m^3\,\delta_{m+n,0},\\ 
  &[J_m,Q^1_r]=-\left(-\frac{m}{2}+r\right)Q^1_{m+r}, \quad [J_m,Q^2_r]=-\left(-\frac{m}{2}+r\right)Q^2_{m+r},\\
    &[J_m,\mathcal{R}_n]=-n\mathcal{R}_{n+m}+C_{jr}^{(0)}\,m^2\,\delta_{m+n,0},\\ &[J_m,\mathcal{S}_n]=-n\mathcal{S}_{n+m},\\
  & [\mathcal{R}_m, Q_r^1]= Q_{m+r}^1,\quad  [\mathcal{R}_m, Q_r^2]=- Q_{m+r}^2, \\
  & \{Q_r^1,Q_s^2\}=P_{r+s}+(r-s)\mathcal{S}_{r+s}+2C_{jp}^{(5)}\,r^2\,\delta_{r+s,0},\quad [P_m,\mathcal{R}_n]=-2nS_{n+m},\\
  &[\mathcal{R}_m,\mathcal{R}_n]=C_{rr}\,m\,\delta_{m+n,0}\ ,\ [\mathcal{R}_m, \mathcal{S}_n]=2C_{jp}^{(5)}m \delta_{m+n,0}\ .
\end{aligned}
\end{equation}
The above is largely in agreement with the results of \cite{Banerjee:2019lrv} except for the $[J_m, \mathcal{R}_n]$ case, where we find a new central term. It is  interesting to understand the source of this central term in the three dimensional asymptotically flat bulk supergravity theory. We have elaborated on the importance of this new central term in the discussion session.

The above analysis gives a complete classification of all possible central extensions to the $W(a,b)$ algebra for arbitrary values of $a$ and $b$. Although, our ansatze \eqref{eq:centralextansatz1}-\eqref{eq:centralextansatz9} was very general, we eventually ended up with only a few non-zero central extensions for certain specific values of $a$ and $b$. Certain central extension such as $z(m,n)$, defined in \eqref{eq:JScentral} and following \eqref{eq:Centralzeqn} did yield non-trivial solutions. However, demanding consistency with subsequent Jacobi indentities led us to conclude $z(m,n)=0$ identically for all values of $a$ and $b$. To summarize our findings, we tabulate the non-trivial central charges obtained for other specific domain of the deformation parameters $a$ and $b$ in the following :\vspace{1em}
\begin{table}[h]
    \centering
    \begin{tabular}{||c|c|c|c|c||}
    \hline
   \textbf{Central extensions}
    &$a=0, b=-1$ & $a=0, b=1$ & $a=0,b=2$ & $a\neq 0, b\neq 0$ \\
    \hline
    $v(m,n)$&$C_{jp}^{(5)}m^3\delta_{m+n,0}$ &0&0&0\\
    \hline
    $f(r,s)$ &$2C_{jp}^{(5)}r^2\delta_{r+s,0}$& $C_{qq}^{(1)}\delta_{r+s,0}$ & $C_{qq}^{(2)}r^2 \delta_{r+s,0}$ & 0\\
    \hline
    $k(m,n)$&$2C_{jp}^{(5)}m\delta_{m+n,0}$&0&$C_{qq}^{(2)}m\delta_{m+n,0}$ &0 \\
    \hline
    \end{tabular}
    \caption{Central Extensions}
    \label{table1}
\end{table}

Note that in the above we have not included the central extension in the $[J_m,J_n]$ commutator which is the usual Virasoro central extension $C_{jj}m^3\delta_{m+n}$ that is present for any values of $a$ and $b$. In addition the central terms in the $[J_m, {\cal R}_n]$ and $[{\cal R}_m, {\cal R}_n]$ commutator also exists for all values of $a$ and $b$ and is given by $y(m,n)=C_{jr}^{(0)}m^2\delta_{m+n,0}$ and $w(m,n)=C_{rr}m\delta_{m+n,0}$ respectively. For all other cases other than the above ones, the central extensions vanish. The above forms of the central extensions along with the commutators given in equations \eqref{eq:centralextansatz1}-\eqref{eq:centralextansatz9} constitute the most generic centrally extended $\mathcal{N}=$2 supersymmetric $W(a,b)$ algebra in presence of $R-$symmetry.

\section{ $\BMSf$ group and \texorpdfstring{$W(a,b;\bar{a},\bar{b})$ algebra}{W(a,b;bar{a},bar{b})}}\label{sec:bms4-intro}

Having established a realization of the extended $W(a,b)$ algebra, we now move on to generalize the above analysis in four spacetime dimensions. Like before, our starting point will be the $\BMSf$ algebra which can be thought of as a special case of the more general $W(a,b; \bar{a},\bar{b})$ algebra \cite{Safari:2019zmc, Safari:2020pje}. 
In the early 1960s, \cite{BMvdB:1962,Sachs:1962} attempted to understand and study the radiation that will be detected by a distant observer. Interestingly, they found that the full set of symmetries for an asymptotically flat spacetime\footnote{There are various equivalent ways in which one can specify asymptotic behaviour of spacetimes. For a detailed exposition, the reader is urged to consult \cite{Frauendiener:2000mk,Ashtekar:2014zfa} and references therein.} is an infinite dimensional group spanned by the so-called supertranslation and superrotations generators which is dubbed the $\BMSf$ group.

The infinite dimensional centerless asymptotic symmetry algebra of four dimensional flat spacetime, conventionally known as the $\BMSf$ algebra \cite{Barnich:2011mi,Safari:2019zmc}, is given by
\begin{align}
[\mathcal{L}_m,\mathcal{L}_n]&=(m-n)\mathcal{L}_{m+n}\notag\\
\left[\bar{\mathcal{L}}_m,\bar{\mathcal{L}}_n\right]&=(m-n)\bar{\mathcal{L}}_{m+n}\notag\\
\left[\mathcal{L}_m,\bar{\mathcal{L}}_n\right]&=0\notag\\
\left[\mathcal{L}_m,T_{p,q}\right]&=\left(\frac{m+1}{2}-p\right)T_{p+m,q}\notag\\
\left[\bar{\mathcal{L}}_m,T_{p,q}\right]&=\left(\frac{m+1}{2}-q\right)T_{p,q+m}\notag\\
\left[T_{p,q},T_{k,l}\right]&=0\label{B}
\end{align}
where the indices $m,n,p,q,k,l \in \mathbb{Z}$. 
The generators $\mathcal{L}_m$ and $\bar{\mathcal{L}}_m$ forming two independent copies of the Witt algebra are known to correspond to \emph{superrotations} while the generators $T_{p,q}$ are known to correspond to \emph{supertranslations}. Following \cite{Caroca:2018obf,Safari:2019zmc}, we briefly state the map between the global sector of $\BMSf$ algebra and Poincar\'{e} algebra.

Denoting the Lorentz generators as $M_{\mu \nu}$ and the translations as $P_{\mu}$, we known that in four spacetime dimensions, they satisfy the algebra
\begin{eqnarray}
\label{eq:PA-1}
\left[M_{\mu \nu},M_{\rho \sigma}\right]&=&i(\eta_{\mu \rho}M_{\nu \sigma}+\eta_{\sigma \mu}M_{\rho \nu}-\eta_{\nu \rho}M_{\mu \sigma}-\eta_{\sigma \nu}M_{\rho \mu})\\
\left[M_{\mu \nu},P_{\sigma}\right]&=&i(\eta_{\sigma \mu}P_{\nu}-\eta_{\sigma \nu}P_{\mu})\\
\label{eq:PA-3}
\left[P_{\mu},P_{\nu}\right]&=&0
\end{eqnarray}
where the indices $\mu, \nu, \rho, \sigma \in \{0,1,2,3\}$ and $\eta_{\mu \nu} \equiv \mbox{diag}(-1,+1,+1,+1)$ is the flat Minkowski metric. We can define the generator of rotations and boosts as
\begin{equation}
J_i=\frac{1}{2}\epsilon_{ijk}M^{jk} \quad \mbox{and} \quad K_i=M^{0i}
\end{equation}
respectively, where $\epsilon_{ijk}$ is the Levi-Civita tensor and the indices $i, j, k \in \{1,2,3\}$. Further, we define the quantities
\begin{eqnarray}
\label{globalBMS4-1}
\mathcal{L}_{\pm 1} &=& iS_1\pm S_2\\
\bar{\mathcal{L}}_{\pm 1}&=& iR_1 \pm R_2\\
\mathcal{L}_0&=&S_3\\
\label{globalBMS4-4}
\bar{\mathcal{L}}_0&=&R_3
\end{eqnarray}
where 
\begin{equation}
R_i=\frac{1}{2}(J_i+iK_i) \quad \mbox{and} \quad S_i=\frac{1}{2}(J_i -iK_i).
\end{equation}
It can be easily verified that the set of operators $\{\mathcal{L}_{\pm 1},\mathcal{L}_0,\bar{\mathcal{L}}_{\pm 1},\bar{\mathcal{L}}_0\}$ satisfies the algebra 
\begin{eqnarray}
\label{ll-comm}
\left[\mathcal{L}_m,\mathcal{L}_n\right]&=&(m-n)\mathcal{L}_{m+n}\\
\label{lbar-lbar-comm}
\left[\bar{\mathcal{L}}_m,\bar{\mathcal{L}}_n\right]&=&(m-n)\bar{\mathcal{L}}_{m+n}\\
\label{l-lbar-comm}
\left[\mathcal{L}_m,\bar{\mathcal{L}}_n\right]&=&0
\end{eqnarray}
for $(m,n) \in \{\pm 1,0\}$ thus showing that the set of operators defined in \eqref{globalBMS4-1}-\eqref{globalBMS4-4} indeed correspond to the global part of the infinite dimensional $\BMSf$ algebra. 

The translation generators $P_{\mu}$ can be mapped to linear combinations of $T_{p,q}$ where $(p,q) \in \{0,1\}$ as follows:
\begin{eqnarray}
\label{eq:TP0-reln}
P^0&=&H=(T_{1,0}-T_{0,1})\\
P^1&=&(-i)(T_{1,1}+T_{0,0})\\
P^2&=&T_{1,1}-T_{0,0}\\
\label{eq:TP3-reln}
P^3&=&T_{1,0}+T_{0,1}
\end{eqnarray}
This demonstrates that appropriate combinations of the global part of $\BMSf$ algebra consisting of the operators $\{\mathcal{L}_{\pm 1},\mathcal{L}_0,\bar{\mathcal{L}}_{\pm 1},\bar{\mathcal{L}}_0,T_{1,0},T_{0,1},T_{0,0},T_{1,1}\}$ can be suitably repackaged to give the Poincar\'{e} algebra. It is noteworthy that the $T_{0,1}, T_{1,0}, T_{0,0}\ \mbox{and}\ T_{1,1}$ gives rise to the translation generators while a certain combination of $(\mathcal{L}_m,\bar{\mathcal{L}}_m)$ where $(m,n) \in \{\pm 1,0\}$ gives rise to the Lorentz generators. In some sense, this gives us further intuition to associate $T_{p,q}$ with \emph{supertranslations} while associating $\mathcal{L}_m$ and $\bar{\mathcal{L}}_m$ with \emph{superrotations}.

Similarly, for four spacetime dimensions it has been proved that $\BMSf$ algebra is not rigid and can be deformed into four parameters family algebra called $W(a,b;\bar{a},\bar{b})$ algebra with commutators as
\begin{equation} 
\begin{split}
 & [\mathcal{L}_{m},\mathcal{L}_{n}]=(m-n)\mathcal{L}_{m+n}, \\
 &[\Bar{\mathcal{L}}_{m},\Bar{\mathcal{L}}_{n}]=(m-n)\Bar{\mathcal{L}}_{m+n},\\
 &[\mathcal{L}_{m},\Bar{\mathcal{L}}_{n}]=0.\\
 &[\mathcal{L}_{m},T_{p,q}]=-(a+bm+p)T_{m+p,q},\\
 &[\Bar{\mathcal{L}}_{n},T_{p,q}]=-(\bar{a}+\bar{b}n+q)T_{p,n+q},\\
 &[T_{p,q},T_{k,l}]=0,
\end{split}
\end{equation}
where $a,b,\bar{a}$ and $\bar{b}$ are arbitrary real parameters. In this way, $\BMSf$ algebra (\ref{B}) can be viewed as $W(-\frac{1}{2},-\frac{1}{2};-\frac{1}{2},-\frac{1}{2})$. This algebra can be viewed as two copies of $W(a,b)$ and $W(\bar{a},\bar{b})$ with the identification of $T_{p,q}$ as a product of supertranslation generators of both the algebras. However, as we see in the next section, this structure does not extend in the supersymmetric extensions of the algebra. Another interesting case is $W(0,0;0,0)$, which represents an infinite dimensional algebra of the symmetries of the near horizon geometry of nonextremal black holes \cite{Donnay:2019zif}. The algebra with $a=b=\bar{a}=\bar{b}=-\frac{1+s}{2}$ for $0<s<1$ describes the asymptotic symmetry algebra of decelerating FLRW spacetime \cite{Enriquez-Rojo:2021blc}. 

\section{Supersymmetric \texorpdfstring{$W(a,b;\bar{a},\bar{b})$\label{sec4} algebra}{W(a,b;bar{a},bar{b})} from $\mathcal{N}=2$ super-$\BMSf$}
In this section we write down a supersymmetric extension of $W(a,b;\bar{a},\bar{b})$ algebra with two supercharges. To get to this,
like the $\BMSt$ algebra, our first goal is to write the supersymmetrized $\BMSf$ algebra. The first attempt towards the construction of a super-$\mathfrak{bms}$ algebra was carried out in \cite{AWADA198652}, which however did not consider superrotation generators. \cite{Avery:2015iix}, further explored asymptotic fermionic charges in ${\cal N}=1$ supergravity on four dimensional asymptotically flat background {\footnote{There also exists a realization of super-$\mathfrak{bms}$ algebra at spatial infinity \cite{Henneaux:2020ekh, Fuentealba:2020aax}}}. \cite{Fotopoulos:2019vac,Fotopoulos:2020bqj} have derived such an algebra by analysing OPEs of appropriate operators of Einstein-Yang-Mills theory at the celestial sphere. However, their convention for indices on the supertranslation generators is different from ours. This changes the index structure that appears in the commutators. We fix the index structure of the supersymmetrized  $\BMSf$ algebra by demanding consistency between its global part and the four dimensional super-Poincar\'{e} algebra which along with \eqref{eq:PA-1}-\eqref{eq:PA-3} now also contains
\begin{equation}
\label{eq:4DsuperPoincare1}
\left\{\cQ_A,\bar{\cQ}_{\dot{B}}\right\}=2(\sigma^{\mu})_{A\dot{B}}P_{\mu}\ \ ,\
\left[M^{\mu \nu},\cQ_A\right]=i(\sigma^{\mu \nu})_A^{\ B}\cQ_B\ \ ,\ [M^{\mu \nu},\bar{\cQ}^{\dot{A}}]=i(\bar{\sigma}^{\mu \nu})^{\dot{A}}_{\ \dot{B}}\bar{\cQ}^{\dot{B}}\ .
\end{equation}
Our starting point in the current context is the algebra stated at \eqref{B}. 
However, as mentioned earlier, we need to determine the index structure once we include the super-current generators which we denote by $Q^i_r$ and $\bar{Q}^i_r$ where $i=1,2$ while $r \in \mathbb{Z} \half$. We begin with the global algebra described in the earlier section \ref{sec:bms4-intro} to determine the indices. Hence, we propose the following ansatz involving the supertranslation and superrotation generators with the fermionic supercurrent generators
\begin{equation}
\label{eq:sBMS4-ansatz}
    \{Q_r^i,\bar{Q}_s^j\}=\delta^{ij}T_{f(r,s), g(r,s)}\ \ ;\ \ [\mathcal{L}_m,Q^i_r]=\alpha(m,r)Q^i_{h(m,r)}\ \ ;\ \ [\bar{\mathcal{L}}_m,\bar{Q}^i_r]=\bar{\alpha}(m,r)\bar{Q}^i_{\bar{h}(m,r)}\ , 
\end{equation}
and all other elements in the super-algebra are zero. We must note here that a priori, it is not necessary that the other possible (anti)-commutators are zero for a four dimensional asymptotically flat supergravity theory, however we consider this simplified deformation for the purpose of this paper. Our approach is pragmatic--we simply want to write a possible supersymmetric extension of $\BMSf$ algebra, such that its global part coincides with the super-Poincar\'{e} algebra, without bothering about the fact if it can be realized from the asymptotic symmetry analysis of a physical supergravity theory. The OPE analysis of \cite{Fotopoulos:2019vac,Fotopoulos:2020bqj} found that the only non-zero commutators in super-$\BMSf$ algebra are the ones mentioned above in \eqref{eq:sBMS4-ansatz}. This motivates us to propose the ansatz as written above. We further make the simplifying assumption that the functions parametrizing the indices $f,g,h,\bar{h}$ are all linear in their arguments \footnote{This structure does not hold for certain symmetry algebras such as the one discussed in \cite{Fuentealba:2021xhn}.}.

We \emph{demand}  the map between supercurrent modes and the fermionic generators of the super-Poincar\'{e} algebra to be
\begin{eqnarray}
	\label{eq:SUSY-generatormapping}
	\cQ^i_{1} \rightarrow Q^i_{\half} \quad &,& \quad \cQ^i_2 \rightarrow Q^i_{\mhalf}\ ,\\
	\bar{\cQ}^i_{\dot{1}} \rightarrow \bar{Q}^i_{\half} \quad &,& \quad \bar{\cQ}^i_{\dot{2}} \rightarrow \bar{Q}^i_{\mhalf}\ .
\end{eqnarray}
Linearity of the indices implies 
\begin{equation}
\label{eq:QQ-indexansatz}
\begin{aligned}
	f(r,s)&=f_0+f_1 r+f_2 s\ ,\\
	g(r,s)&=g_0+g_1 r+g_2 s\ .
\end{aligned}
\end{equation} 
where $f_i$ and $g_i$ are constants. Using \eqref{eq:4DsuperPoincare1}, along with \eqref{eq:TP0-reln}-\eqref{eq:TP3-reln}, we must have
\begin{equation}
\begin{aligned}
	&\{\cQ^i_1, \bar{\cQ}^j_{\dot{1}}\} =-2(P_0-P_3)\delta^{ij}=4T_{1,0}\delta^{ij}\ ,\ \{\cQ^i_1, \bar{\cQ}^j_{\dot{2}}\}=2(P_1-iP_2)\delta^{ij}=-4iT_{1,1}\delta^{ij}\ ,\\
	&\{\cQ^i_2, \bar{\cQ}^j_{\dot{1}}\}=2(P_1+iP_2)\delta^{ij}=-4iT_{0,0}\delta^{ij}\ ,\ \{\cQ^i_2, \bar{\cQ}^j_{\dot{2}}\}=-2(P_0+P_3)\delta^{ij}=-4T_{0,1}\delta^{ij}\ .
\end{aligned}
\end{equation}
Thus, the mapping \eqref{eq:SUSY-generatormapping} requires the functions $f(r,s)$ and $g(r,s)$ to satisfy
\begin{equation}
\label{eq:poincarealgfg}
\begin{aligned}
	&f\left(\half,\half\right)=1\ , f\left(\half,\mhalf\right)=1\ , f\left(\mhalf,\half\right)=0\ , f\left(\mhalf,\mhalf\right)=0\ ,\\
	&g\left(\half,\half\right)=0\ , g\left(\half,\mhalf\right)=1\ , g\left(\mhalf,\half\right)=0\ , g\left(\mhalf,\mhalf\right)=1\ .
\end{aligned}
\end{equation}
Thus, we need to solve for the six unknowns $f_i, g_i (i=0,1,2)$ appearing in \eqref{eq:QQ-indexansatz} from the above eight equations. As it turns out, (detailed in appendix \ref{app:QQbarstructure}) there does exist a consistent solution to the above system given by
\begin{equation}
\label{eq:fgfullsolution}
	f_0=g_0=\half, f_1=-g_2=1, f_2=g_1=0\ .
\end{equation}
Thus, we eventually recover
\begin{equation}
	\{Q_r^i,\bar{Q}_s^j\}=\delta^{ij}T_{r\half, -s\half}\ ,\ \{Q^i_r,Q^j_s\}=\{\bar{Q}^i_r,\bar{Q}^j_s\}=0
\end{equation}
The exact map \eqref{eq:SUSY-generatormapping} can be seen to be
\begin{equation}
    \cQ^i_1=2Q^i_{\half}\ ,\ \bar{\cQ}^i_{\dot{1}}=2\bar{Q}^i_{\half}\ ,\ \cQ_2^i=-2iQ^i_{\mhalf}\ ,\ \bar{\cQ}^i_{\dot{2}}=-2i\bar{Q}^i_{\mhalf}\ .
\end{equation}
Using the above map, along with the map described in Sec. \ref{sec:bms4-intro}, we get,
\begin{equation}
\begin{aligned}
    &[\mathcal{L}_{-1},Q^i_{\half}]=-Q^i_{\mhalf}\ ,\ [\mathcal{L}_0,Q^i_{\half}]=-\frac{1}{2}Q^i_{\half}\ ,\ [\mathcal{L}_{+1},Q^i_{\half}]=0\ , \\
    &[\mathcal{L}_{-1},Q^i_{\mhalf}]=0\ ,\ [\mathcal{L}_{0},Q^i_{\mhalf}]=\frac{1}{2}Q^i_{\mhalf}\ ,\ [\mathcal{L}_{+1},Q^i_{\mhalf}]=Q^i_{\half}\ .
\end{aligned}
\end{equation}
Further assuming linearity of the structure constant $\alpha(m,r)$ and $h(m,r)$ appearing in \eqref{eq:sBMS4-ansatz}, we see that the above global sector is consistent provided one has
\begin{equation}
    [\mathcal{L}_m, Q^i_r]= \left(\frac{m}{2}-r\right)Q^i_{m+r}\ .
\end{equation}
An identical exercise on the ``barred" sector first leads us to the relations
\begin{equation}
\begin{aligned}
    &[\bar{\mathcal{L}}_{-1},\bar{Q}^i_{\half}]=0\ ,\ [\bar{\mathcal{L}}_0,\bar{Q}^i_{\half}]=\frac{1}{2}\bar{Q}^i_{\half}\ ,\ [\bar{\mathcal{L}}_{+1},\bar{Q}^i_{\half}]=\bar{Q}^i_{\mhalf}\ , \\
    &[\bar{\mathcal{L}}_{-1},\bar{Q}^i_{\mhalf}]=-\bar{Q}^i_{\half}\ ,\ [\bar{\mathcal{L}}_{0},\bar{Q}^i_{\mhalf}]=-\frac{1}{2}\bar{Q}^i_{\mhalf}\ ,\ [\bar{\mathcal{L}}_{+1},\bar{Q}^i_{\mhalf}]=0\ .
\end{aligned}
\end{equation}
This shows that we must have
\begin{equation}
    [\bar{\mathcal{L}}_m, \bar{Q}^j_{s}]=\left(\frac{m}{2}+s\right)\bar{Q}^j_{-m+s}\ .
\end{equation}
Thus, a particular realization of the $\mathcal{N}=2$ super-$\BMSf$ algebra can be written as
\begin{equation} 
\label{eq:deformed-BMS4}
\begin{split}
 & [\mathcal{L}_{m},\mathcal{L}_{n}]=(m-n)\mathcal{L}_{m+n}, \\
 &[\bar{\mathcal{L}}_{m},\bar{\mathcal{L}}_{n}]=(m-n)\bar{\mathcal{L}}_{m+n},\\
 &[\mathcal{L}_{m},T_{p,q}]=\left(\frac{m+1}{2}-p\right)T_{m+p,q},\\
 &[\bar{\mathcal{L}}_{m},T_{p,q}]=\left(\frac{m+1}{2}-q\right)T_{p,m+q},\\
 &\{Q^i_r,\bar{Q}^j_s\}=\delta^{ij}T_{r+1/2,-s+1/2},\\
 &[\mathcal{L}_m,Q^i_r]=\left(\frac{m}{2}-r\right)Q^i_{m+r},\\
 &[\bar{\mathcal{L}}_m,\bar{Q}^i_r]=\left(\frac{m}{2}+r\right)\bar{Q}^i_{-m+r}\ .
\end{split}
\end{equation}
while the other (anti)-commutators are identically zero. One can easily check that all the Jacobi identities are satisfied for the above written algebra. Now, that we have fixed the indices in the super-$\BMSf$ algebra, we assume that deformations do not change that and thus will carry over to the $W(a,b;\bar{a},\bar{b})$ algebra. Thus, we now propose the following supersymmetrized $W(a,b;\bar{a},\bar{b})$ algebra
\begin{equation}
\label{eq:superW4-ansatz}
\begin{split}
 & [\mathcal{L}_{m},\mathcal{L}_{n}]=(m-n)\mathcal{L}_{m+n}, \\
 &[\bar{\mathcal{L}}_{m},\bar{\mathcal{L}}_{n}]=(m-n)\bar{\mathcal{L}}_{m+n},\\
 &[\mathcal{L}_{m},T_{p,q}]=-\left(a+bm+p\right)T_{m+p,q},\\
 &[\bar{\mathcal{L}}_{m},T_{p,q}]=-\left(\bar{a}+\bar{b}m+q\right)T_{p,m+q},\\
 &\{Q^i_r,\bar{Q}^j_s\}=\delta^{ij}T_{r+1/2,-s+1/2},\\
 &[\mathcal{L}_m,Q^i_r]=\alpha(m,r)Q^i_{m+r},\\
 &[\bar{\mathcal{L}}_m,\bar{Q}^i_r]=\bar{\alpha}(m,r)\bar{Q}^i_{-m+r}\ .
\end{split}
\end{equation}
Strictly speaking, the most general extension would require us to choose all such as $[\mathcal{L}_m,\bar{Q}^i_r], [\bar{\mathcal{L}}_m,{Q}^i_r]$ to be non-zero and demand consistency of Jacobi identities. However we have chosen the above as a possible extension to verify if it satisfies all Jacobi identities and thus making it a consistent graded Lie algebra. The Jacobi identity for $Q^i_r, \bar{Q}^j_s$ and $\mathcal{L}_m$ is given by
\begin{equation}
    [\{Q_r^i,\bar{Q}_s^j\},\mathcal{L}_m ]= \{Q_r^i,[\bar{Q}_s^j,\mathcal{L}_m]\}+\{\bar{Q}_s^j,[Q_r^i,\mathcal{L}_m]\}
\end{equation}
Using \eqref{eq:superW4-ansatz}, we get,
\begin{equation}
    \alpha(m,r)=-\left(a+bm+r+\frac{1}{2}\right)\ .
\end{equation}
The Jacobi identities of $Q^i_r, \bar{Q}^j_s$ and $\bar{\mathcal{L}}_m$ similarly gives
\begin{equation}
    \bar{\alpha}(m,r)=-\left(\bar{a}+\bar{b}m-r+\frac{1}{2}\right)\ .
\end{equation}
Thus, a ${\mathcal N}=2$ supersymmetric $W(a,b;\bar{a},\bar{b})$ algebra is given by
\begin{equation}
\label{eq:susyW-4d}
\begin{split}
 & [\mathcal{L}_{m},\mathcal{L}_{n}]=(m-n)\mathcal{L}_{m+n}, \\
 &[\bar{\mathcal{L}}_{m},\bar{\mathcal{L}}_{n}]=(m-n)\bar{\mathcal{L}}_{m+n},\\
 &[\mathcal{L}_{m},T_{p,q}]=-\left(a+bm+p\right)T_{m+p,q},\\
 &[\bar{\mathcal{L}}_{m},T_{p,q}]=-\left(\bar{a}+\bar{b}m+q\right)T_{p,m+q},\\
 &\{Q^i_r,\bar{Q}^j_s\}=\delta^{ij}T_{r+1/2,-s+1/2},\\
 &[\mathcal{L}_m,Q^i_r]=-\left(a+bm+r+\frac{1}{2}\right)Q^i_{m+r},\\
 &[\bar{\mathcal{L}}_m,\bar{Q}^i_r]=-\left(\bar{a}+\bar{b}m-r+\frac{1}{2}\right)\bar{Q}^i_{-m+r}\ ,
\end{split}
\end{equation}
and all other commutators are zero. It satisfies all the Jacobi identities. The above algebra is a simple extension of bosonic $W(a,b;\bar{a},\bar{b})$ algebra with two supercurrent generators, as it does not contain any rotation among the supercurrent generators. In the context of super-Poincar\'{e} algebra, the $R$-charge rotates the SUSY generators amongst themselves. To be mathematically precise, the $R-$extension is the largest subgroup of the automorphism group of the supersymmetry algebra which commutes with the Lorentz group. In the next section we perform the  $R$-extension of the above ${\mathcal N}=2$ supersymmetric $W(a,b;\bar{a},\bar{b})$ algebra, by further introducing the R-charge generators to rotate the two supercurrent generators. 

\section{$R$-extended supersymmetric \texorpdfstring{$W(a,b;\bar{a},\bar{b})$ algebra}{W(a,b;bar{a},bar{b})}}\label{sec5}
Before going to $R$-extension of the $W(a,b;\bar{a},\bar{b})$ algebra, let us briefly recall some important points about super-Poincare algebras in presence of $R-$symmetry. 
As discussed in \cite{Hori:2003ic} generic $\mathcal{N}=2$ super-Poincare algebra contains two species of $R$-symmetry generators--vectorial and axial which act on the supercharges as
\begin{eqnarray}
\label{eq:Rsymmglobal1}
[Q^i_{\pm \frac{1}{2}},R_0]=Q^i_{\pm \frac{1}{2}}\ &,&\ [\bar{Q}^i_{\pm \frac{1}{2}},R_0]=-\bar{Q}^i_{\pm \frac{1}{2}}\ ,\\
\label{eq:Rsymmglobal2}
{[Q^i_{\pm \frac{1}{2}},\bar{R}_0]}= \pm Q^i_{\pm \frac{1}{2}}\ &,&\ [\bar{Q}^i_{\pm \frac{1}{2}}, \bar{R}_0]=\mp \bar{Q}^i_{\pm \frac{1}{2}}
\end{eqnarray}
As a matter of fact theories with extended supersymmetries are extremely rich precisely due to the presence of these two kinds of supercharges. These $R$-symmetries can in fact be used as a powerful tool to define various \emph{twistings} in a theory (A-type or B-type) resulting in what is known as Topological Field Theories whose correlators happen to be independent of the background metric. Although conventionally theories with either one kind of $R$-symmetry are considered, in general the full theory does contain both kinds of R-symmetries. In the current context since we are specifically interested in $\mathcal{N}=2$ SUSY, the $R$-symmetry generator in fact generates the group $U(1)_V \times U(1)_A$. Following \cite{Hori:2003ic}, we will associate $R_0$ with the vectorial $R$-symmetry while $\bar{R}_0$ will be associated as the axial $R$-symmetry. Another important aspect of super-Poincare algebras is the fact that the $R-$charges commute with the bosonic Poincare generators. We utilize these facts in our constructions. To be precise, in the following we demand that our $R$-extended $W(a,b;\bar{a},\bar{b})$ algebra must have vanishing commutators of the $R-$charge generators with other bosonic generators in the global sector for the deformation $W(-1/2,-1/2;-1/2,-1/2)$\footnote{$W(-1/2,-1/2;-1/2,-1/2)$ is the $\BMSf$ algebra whose global sector coincides with the $\mathcal{N}=2$ super-Poincare algebra.}

In order to make the extension of $R$-charges in the context of $W(a,b;\bar{a},\bar{b})$ algebra, we will start with a general ansatz,
\begin{equation}
\label{eq:BMSf-RSansatz}
\begin{split}
&[Q_r^i,R_n]=\beta^{(i)}(r,n) Q^i_{\theta_1(r,n)}\ ,\ [\bar{Q}_r^i,R_n]=\bar{\beta}^{(i)}(r,n) \bar{Q}^i_{\bar{\theta}_1(r,n)}\ ,\\
&{[Q_r^i, \bar{R}_n]}=\kappa^{(i)}(r,n)Q^i_{\theta_2(r,n)}\ ,\ [\bar{Q}_r^i, \bar{R}_n]=\bar{\kappa}^{(i)}(r,n)\bar{Q}^i_{\bar{\theta}_2(r,n)}\ .
\end{split}
\end{equation}
It must be noted that in the above ansatz, although $i$ is a repeated index on the RHS, there is no sum over $i$. Although we will be primarily interested in $\mathcal{N}=2$ but the discussion in this section is valid for arbitrary value of $\mathcal{N}$. Thus we consider $i=1,2,..,\mathcal{N}$. Furthermore, since we are working in a very general setting, we assume that $\beta^{i}$ is different for $i=1,2,..,\mathcal{N}$. This is of course not the most general extension one can think of but rather a simpler starting point which is also consistent with the global sub-sector \eqref{eq:Rsymmglobal1}-\eqref{eq:Rsymmglobal2}. 

In the following analysis, we will assume that the indices appearing in the above ansatz are linear in its arguments, which is a basic feature of most algebras. Consistency with the global subsector i.e. \eqref{eq:Rsymmglobal1}-\eqref{eq:Rsymmglobal2} fixes the form of the indices as
\begin{equation}
\label{eq:BMSQRfixed}
\begin{split}
&[Q_r^i,R_n]=\beta^{(i)}(r,n) Q^i_{cn+r}\ ,\ [\bar{Q}_r^i,R_n]=\bar{\beta}^{(i)}(r,n) \bar{Q}^i_{\bar{c}n+r}\\
&{[Q_r^i, \bar{R}_n]}=\kappa^{(i)}(r,n)Q^i_{kn+r}\ ,\ [\bar{Q}_r^i, \bar{R}_n]=\bar{\kappa}^{(i)}(r,n)\bar{Q}^i_{\bar{k}n+r}\ .
\end{split}
\end{equation}
where $c,\bar{c},k$ and $\bar{k}$ are integers and the structure constants are non-zero at least in the global subsector i.e. when $r= \pm \frac{1}{2}$ and $n=0$. We have to find their form away from the global sector. Subsequently, we will concentrate on the commutators $[R_n, T_{p,q}]$ and $[\bar{R}_n, T_{p,q}]$. Consider the Jacobi identity for the operators $Q^i_r, \bar{Q}^j_s$ and $R_n$ which gives us
\begin{equation}
    [\{Q_r^i,\bar{Q}_s^j\},R_n ]= \{Q_r^i,[\bar{Q}_s^j,R_n]\}+\{\bar{Q}_s^j,[Q_r^i,R_n]\}\ .
\end{equation}
Using \eqref{eq:BMSQRfixed}, in the above, we get,
\begin{equation}
  \delta^{ij}{[T_{p,q},R_n]}=\delta^{ij}\bigg[\bar{\beta}^{(j)} \left(-q\half,n\right)T_{p,-\bar{c}n+q}+\beta^{(i)} \left(p\mhalf,n\right)T_{cn+p,q}\bigg]. \nonumber
\end{equation}
In the above equation $i,j$ are free indices. In particular for $\mathcal{N}=2$ SUSY , we see that,
\begin{equation}
\label{eq:TR-commutator}
\begin{aligned}
    {[T_{p,q},R_n]}&=\bar{\beta}^{(1)}\left(-q\half,n\right)T_{p,-\bar{c}n+q}+\beta^{(1)}\left(p\mhalf,n\right)T_{cn+p,q}\\
    &=\bar{\beta}^{(2)}\left(-q\half,n\right)T_{p,-\bar{c}n+q}+\beta^{(2)}\left(p\mhalf,n\right)T_{cn+p,q}\ .
\end{aligned}
\end{equation}
which also implies $\beta^{(1)}(r,n)=\beta^{(2)}(r,n)$ and $\bar{\beta}^{(1)}(r,n)=\bar{\beta}^{(2)}(r,n)$. An identical exercise with $Q^i_r, \bar{Q}^j_s$ and $\bar{R}_n$ gives
\begin{equation}
\begin{aligned}
    {[T_{p,q},\bar{R}_n]}&=\bar{\kappa}^{(1)}\left(-q\half,n\right)T_{p,-\bar{k}n+q}+\kappa^{(1)}\left(p\mhalf,n\right)T_{kn+p,q}\\
    &=\bar{\kappa}^{(2)}\left(-q\half,n\right)T_{p,-\bar{k}n+q}+\kappa^{(2)}\left(p\mhalf,n\right)T_{kn+p,q}\ .
\end{aligned}
\end{equation}
leading to the condition that $\kappa^{(1)}(r,n)=\kappa^{(2)}(r,n)$ and $\bar{\kappa}^{(1)}(r,n)=\bar{\kappa}^{(2)}(r,n)$. Given the relation between the structure constants for $i=1,2$, we see that the index is extraneous and hence we will simply be dropping it from our notation subsequently. Thus, we may write more simply
\begin{equation}
\begin{aligned}\label{eq:QRT-simplified}
&[Q_r^i,R_n]=\beta(r,n) Q^i_{cn+r}\ ,\ [\bar{Q}_r^i,R_n]=\bar{\beta}(r,n) \bar{Q}^i_{\bar{c}n+r}\ ,\\
&{[Q_r^i, \bar{R}_n]}=\kappa(r,n)Q^i_{kn+r}\ ,\ [\bar{Q}_r^i, \bar{R}_n]=\bar{\kappa}(r,n)\bar{Q}^i_{\bar{k}n+r}\ ,\\
&{[T_{p,q},R_n]}=\bar{\beta}\left(-q\half,n\right)T_{p,-\bar{c}n+q}+\beta \left(p\mhalf,n\right)T_{cn+p,q}\ ,\\
&{[T_{p,q},\bar{R}_n]}=\bar{\kappa}\left(-q\half,n\right)T_{p,-\bar{k}n+q}+\kappa\left(p\mhalf,n\right)T_{kn+p,q}\ .
\end{aligned}
\end{equation}

Note that the above commutation relations also ensure that the Jacobi identities between $T_{p,q}, T_{m,n}, R_l$ as well as $T_{p,q}, Q^i_r, R_n$ (and it's corresponding counterparts with $Q^i_r$ replaced with $\bar{Q}^j_s$ and $R_n$ replaced with $\bar{R}_n$) are also satisfied. Now, that we have identified $R_n$ and $\bar{R}_n$ to the vectorial and axial $R$-supercurrents, we make a further assumption that 
\begin{equation}
    [R_n,\bar{R}_m]=0\ .
\end{equation}
The above assumptions applied to the Jacobi identity of $R_n, \bar{R}_m$ and $Q^i_r$ leads to a relation between the structure constants 
\begin{equation}
\label{eq:RRbarQJacobi}
    \kappa(cn+r,m)\beta(r,n)=\kappa(r,m)\beta(km+r,n)\ ,
\end{equation}
and analogously the Jacobi identity of $R_n, \bar{R}_m$ and $\bar{Q}^i_r$ gives rise to,
\begin{equation}
\label{eq:RRbarQbarJacobi}
    \bar{\kappa}(\bar{c}n+r,m)\bar{\beta}(r,n)=\bar{\kappa}(r,m)\bar{\beta}(\bar{k}m+r,n)\ .
\end{equation}
The above two relations seems to put certain constraints on the free parameters $c,\bar{c},k$ and $\bar{k}$. However, we will explore this subsequently.

Finally, we need to fix the algebra between the $R$-supercurrents and the superrotations $\mathcal{L}_m$ and $\bar{\mathcal{L}}_m$. For this purpose we consider a set of Jacobi identities and fixing the form of the commutators between the $R$-charge supercurrents and the superrotation generators. 

\begin{enumerate}
\item \textbf{Jacobi identity for $\mathcal{L}_m, Q^i_r, R_n$ and $\mathcal{L}_m, \bar{Q}^j_s, R_n$}\\
We start with the Jacobi identity for $\mathcal{L}_m, Q^i_r$ and $R_n$ which is given by
\begin{equation}
    [[\mathcal{L}_m, Q^i_r],R_n]+[[Q^i_r,R_n],\mathcal{L}_m]+[[R_n,\mathcal{L}_m],Q^i_r]=0\ .
\end{equation}
Using \eqref{eq:susyW-4d} and \eqref{eq:QRT-simplified}, we can simplify the above equation to obtain
\begin{equation}
\label{eq:RLQJacobi}
    [[R_n,\mathcal{L}_m],Q^i_r]+[\alpha(m,r)\beta(m+r,n)-\beta(r,n)\alpha(m,cn+r)]Q^i_{cn+m+r}=0\ .
\end{equation}
Clearly, looking at the above, on very general grounds, one can schematically write,
\begin{equation}
\label{eq:possibleRL}
     [R_n, \mathcal{L}_m] = w_1(n,m)\mathcal{L}_{t_1(n,m)}+h_1(n,m)R_{u_1(n,m)}+\bar{h}_1(n,m)\bar{R}_{v_1(n,m)}\ .
\end{equation}
The Jacobi identity for $\mathcal{L}_m, \bar{Q}^j_s$ and $R_n$ along with \eqref{eq:susyW-4d} and \eqref{eq:BMSQRfixed} leads to
\begin{equation}
\label{eq:LQbarRJacobi}
    [[R_n, \mathcal{L}_m], \bar{Q}^j_s]=0\ .
\end{equation}
Now, if we consider the $[R_n,\mathcal{L}_m]$ to be of the form as \eqref{eq:possibleRL}, we easily see that the part $[R_{u(n,m)},\bar{Q}^j_s]$ and $[\bar{R}_{v(n,m)},\bar{Q}^j_s]$ will be generically non-zero \emph{individually} for arbitrary values of $n,m$ and $s$ however a linear combination with specific forms of $h_1$ and $\bar{h}_1$ might presumably ensure that the expression vanishes. Note that \eqref{eq:possibleRL} has certain features which puts it in stark contrast to its 3 dimensional $W(a,b)$ analog. Firstly, the first term appearing on the RHS in the above equation has no analog for the $W(a,b)$ algebra as stated explicitly in \eqref{SUSY-W(a,b)}. We can also see, commutators similar to the $W(a,b)$ algebra i.e. 
\begin{equation}
    [R_n, \mathcal{L}_m]\sim R_{u(n,m)}\ \text{and}\ [\bar{R}_n, \mathcal{L}_m]\sim \bar{R}_{v(n,m)}
\end{equation}
is clearly inconsistent with \eqref{eq:LQbarRJacobi} for arbitrary values of the indices $m$ and $n$. 

\item \textbf{Jacobi identity for $\bar{\mathcal{L}}_m, \bar{Q}^j_s, R_n$ and $\bar{\mathcal{L}}_m, Q^i_r$ and $R_n$}\\
The Jacobi identity of $\bar{\mathcal{L}}_m, \bar{Q}^j_s$ and $R_n$ leads us to
\begin{equation}
\label{eq:LbarQbarRJacobi}
\begin{aligned}
    &[[R_n, \bar{\mathcal{L}}_m],\bar{Q}^j_s]\\
    &+\left(\bar{\beta}(s,n)\left(\bar{a}+\bar{b}m-\bar{c}n-s+\frac{1}{2}\right)-\bar{\beta}(-m+s,n)\left(\bar{a}+\bar{b}m-s+\frac{1}{2}\right)\right)\bar{Q}^j_{\bar{c}n+s-m}=0\ .
\end{aligned}
\end{equation}
The above along with the Jacobi identity for $\bar{\mathcal{L}}_m, Q^i_r$ which simplifies to
\begin{equation}
    [[R_n, \bar{\mathcal{L}}_m],Q^i_r]=0\ ,
\end{equation}
suggests of a relation of the form
\begin{equation}
\label{eq:RLbar-ansatz}
    [R_n, \bar{\mathcal{L}}_m] =w_2(n,m)\bar{\mathcal{L}}_{t_2(n,m)}+h_2(n,m)R_{u_2(n,m)}+\bar{h}_2(n,m)\bar{R}_{v_2(n,m)}\ .
    \end{equation}
\item \textbf{Jacobi identity for $\mathcal{L}_m, Q^i_r, \bar{R}_n$ and $\mathcal{L}_m, \bar{Q}^j_s, \bar{R}_n$}\\
The Jacobi identity for $\mathcal{L}_m, Q^i_r$ and $\bar{R}_n$ leads to
\begin{equation}
\label{eq:LQRbarJacobi}
\begin{aligned}
    &[[\bar{R}_m,\mathcal{L}_m],Q^i_r]+\\
    &\left(\kappa(r,n)\left(a+bm+kn+r+\frac{1}{2}\right)-\kappa(m+r,n)\left(a+bm+r+\frac{1}{2}\right)\right)Q^i_{kn+m+r}=0\ ,
\end{aligned}
\end{equation}
while the Jacobi identity for $\mathcal{L}_m, \bar{Q}^j_s$ and $\bar{R}_n$ gives
\begin{equation}
    [[\bar{R}_n, \mathcal{L}_m],\bar{Q}^j_{s}]=0\ .
\end{equation}
This leads us to propose
\begin{equation}
\label{eq:RbarL-ansatz}
    [\bar{R}_n, \mathcal{L}_m]=w_3(n,m)\mathcal{L}_{t_3(n,m)}+h_3(n,m)R_{u_3(n,m)}+\bar{h}_3(n,m)\bar{R}_{v_3(n,m)}\ .
\end{equation}
\item \textbf{Jacobi identity for $\bar{\mathcal{L}}_m, Q^i_r, \bar{R}_n$ and $\bar{\mathcal{L}}_m, \bar{Q}^j_s, \bar{R}_n$}\\
The Jacobi identity for $\bar{\mathcal{L}}_m, Q^i_r, \bar{R}_n$ simplifies to
\begin{equation}
    [[\bar{R}_n,\bar{\mathcal{L}}_m],Q^i_r]=0\ 
\end{equation}
and the $\bar{\mathcal{L}}_m, \bar{Q}^j_s, \bar{R}_n$ Jacobi identity leads to
\begin{equation}
\label{eq:LbarQbarRbarJacobi}
\begin{aligned}
    &[[\bar{R}_n,\bar{\mathcal{L}}_m],\bar{Q}^j_s]\\
    &+\left(\bar{\kappa}(s,n)\left(\bar{a}+\bar{b}m-\bar{k}n-s+\frac{1}{2}\right)-\bar{\kappa}(-m+s,n)\left(\bar{a}+\bar{b}m-s+\frac{1}{2}\right)\right)\bar{Q}^j_{\bar{k}n+s-m}=0\ .
\end{aligned}
\end{equation}
The above two equations lead us to the ansatz
\begin{equation}
\label{eq:RbarLbar-ansatz}
    {[\bar{R}_n,\bar{\mathcal{L}}_m]}=w_4(n,m)\bar{\mathcal{L}}_{t_4(n,m)}+h_4(n,m)R_{u_4(n,m)}+\bar{h}_4(n,m)\bar{R}_{v_4(n,m)}\ .
\end{equation}
\item \textbf{Jacobi identity for $\mathcal{L}_m, T_{p,q}, R_n$ and $\bar{\mathcal{L}}_m, T_{p,q}, R_n$}\\
The Jacobi identity for the operators $\mathcal{L}_m, T_{p,q}$ and $R_n$ is given by
\begin{equation}
    [[\mathcal{L}_m, T_{p,q}],R_n]+[[T_{p,q},R_n],\mathcal{L}_m]+[[R_n,\mathcal{L}_m],T_{p,q}]=0\ ,
\end{equation}
which upon using \eqref{eq:susyW-4d} and \eqref{eq:QRT-simplified} leads us to the relation 
\begin{equation}
\label{eq:LTRJacobi}
\begin{aligned}
    &[[R_n, \mathcal{L}_m],T_{p,q}]\\
    &+\left((a+bm+cn+p)\beta \left(p-\frac{1}{2},n\right)-(a+bm+p)\beta \left(m+p-\frac{1}{2},n\right)\right)T_{cn+m+p,q}=0\ ,
\end{aligned}
\end{equation}
which is consistent with the ansatz \eqref{eq:possibleRL}. Further, the Jacobi identity for $\bar{\mathcal{L}}_m, T_{p,q}, R_n$ gives
\begin{equation}
\label{eq:RLbarTJacobi}
\begin{aligned}
    &[[R_n, \bar{\mathcal{L}}_m],T_{p,q}]\\
    &+\left(\bar{\beta}\left(-q+\frac{1}{2},n\right)(\bar{a}+\bar{b}m-\bar{c}n+q)-\bar{\beta}\left(-m-q+\frac{1}{2},n\right)(\bar{a}+\bar{b}m+q)\right)T_{p,-\bar{c}n+m+q}=0\ ,
\end{aligned}
\end{equation}
which is consistent with the ansatz \eqref{eq:RLbar-ansatz}.

\item \textbf{Jacobi identity for $\mathcal{L}_m, T_{p,q}, \bar{R}_n$ and $\bar{\mathcal{L}}_m, T_{p,q}, \bar{R}_n$}\\
The Jacobi identity for $\mathcal{L}_m, T_{p,q}, \bar{R}_n$ leads to
\begin{equation}
\label{eq:LTRbarJacobi}
    \begin{aligned}
         &[[\bar{R}_n,\mathcal{L}_m],T_{p,q}]\\
         &+\left({\kappa}\left(p-\frac{1}{2},n\right)(a+bm+kn+p)-\kappa\left(m+p-\frac{1}{2},n\right)(a+bm+p)\right)T_{kn+m+p,q}=0\ ,
    \end{aligned}
\end{equation}
while for the other tuple, namely $\bar{\mathcal{L}}_m, T_{p,q}, \bar{R}_n$, we have,
\begin{equation}
\label{eq:LbarTRbarJacobi}
    \begin{aligned}
         &[[\bar{R}_n,\bar{\mathcal{L}}_m],T_{p,q}]\\
         &+\left(\bar{\kappa}\left(-q+\frac{1}{2},n\right)(\bar{a}+\bar{b}m-\bar{k}n+q)-\bar{\kappa}\left(-m-q+\frac{1}{2},n\right)(\bar{a}+\bar{b}m+q)\right)T_{p,-\bar{k}n+m+q}=0\ .
    \end{aligned}
\end{equation}
One can easily see that the above two equations are indeed consistent with the proposed ansatze \eqref{eq:RLbar-ansatz} and \eqref{eq:RbarLbar-ansatz} respectively.
\end{enumerate}
Armed with a series of ansatze and a number of Jacobi identities, we enumerate below the index structure and which equations they follows from.
\begin{itemize}
    \item Plugging ansatz \eqref{eq:possibleRL} in \eqref{eq:RLQJacobi}, we get,
    \begin{equation}
    \label{eq:simplifiedLQR}
    \begin{aligned}
        -w_1(n,m)\alpha(t_1,r)Q^i_{t_1+r}&-h_1(n,m)\beta(r,u_1)Q^i_{cu_1+r}-\bar{h}_1(n,m)\kappa(r,v_1)Q^i_{kv_1+r}\\
        &+[\alpha(m,r)\beta(m+r,n)-\beta(r,n)\alpha(m,cn+r)]Q^i_{cn+m+r}=0\ .
    \end{aligned}
    \end{equation}
    The above equation must be satisfied for arbitrary permissible values of $m,n$ and $r$. Also, by definition, $\beta$ and $\kappa$ cannot be identically zero. This implies,
    \begin{eqnarray}
    u_1(n,m)&=&n+ \frac{m}{c} \nonumber\\
    v_1(n,m)&=&\frac{c}{k}n+\frac{m}{k} \nonumber\ .
    \end{eqnarray}
    Since clearly the indices must be integers, we see both $k$ and $c$ must divide every integer which naturally implies that both can takes values $\pm 1$. Using \eqref{eq:LQbarRJacobi}, we further get,
    \begin{equation}
    \label{eq:xidefn}
        \frac{\bar{c}}{c}=\frac{\bar{k}}{k} =\xi\ (\text{say})\ .
    \end{equation}
    where $\xi$ is some real number. Further looking at \eqref{eq:simplifiedLQR}, we can have either $w_1$ to be identically zero or $t_1= cn+m$. Further, \eqref{eq:LTRJacobi} gives
    \begin{equation}\label{eq:RLTJacobiIdentity}
        \begin{aligned}
             &\left(-w_1(n,m)(a+bcn+bm+p)-h_1(n,m)\beta\left(p-\frac{1}{2},n+\frac{m}{c}\right) \right.\\
             &\left.-\bar{h}_1(n,m)\kappa\left(p-\frac{1}{2},\frac{c}{k}n+\frac{m}{k} \right)+(a+bm+cn+p)\beta \left(p-\frac{1}{2},n\right)\right.\\
             &\left.-(a+bm+p)\beta \left(m+p-\frac{1}{2},n\right)\right)T_{cn+m+p,q}\\
             &+\left(-h_1(n,m)\bar{\beta}\left(-q+\frac{1}{2},n+\frac{m}{c}\right)-\bar{h}_1(n,m)\bar{\kappa}\left(-q+\frac{1}{2},\frac{c}{k}n+\frac{m}{k} \right)\right)T_{p,-\bar{c}n-\xi m+q}=0\ .
        \end{aligned}
    \end{equation}
    In the above expression, each of the coefficients of the $T_{cn+m+p,q}$ and $T_{p,-\bar{c}n-\xi m+q}$ has to vanish individually.\\

\item Ansatz \eqref{eq:RLbar-ansatz} along with \eqref{eq:LbarQbarRJacobi} 
    implies $t_2(n,m)=-\bar{c}n+m$ provided $w_2$ does not vanish identically. Along similar arguments as before, we also conclude
    \begin{eqnarray}
    u_2&=&n- \frac{m}{\bar{c}} \nonumber\\
    v_2&=&\frac{\bar{c}}{\bar{k}}n-\frac{m}{\bar{k}}. \nonumber
    \end{eqnarray}
    Again, since both $u_2$ and $v_2$ must be integers for all values of $m$ and $n$, this leads us to conclude that $\bar{c}$ and $\bar{k}$ can only take values $\pm 1$. Thus, $\xi$ appearing in \eqref{eq:xidefn} can only be $\pm 1$. Now, since $c^2=k^2=\bar{c}^2=\bar{k}^2=\xi^2=1$, we can rewrite 
    \begin{equation}
        u_1= n+mc\ ,\ v_1=kcn+km\ ,\ u_2= n-\xi cm\ ,\ v_2= kcn-\xi km\ .
    \end{equation}
    Further, using \eqref{eq:RLbarTJacobi}, we get, 
    \begin{equation}\label{eq:RLbarTJacobisimp}
    \begin{aligned}    
        &-w_2(n,m)(\bar{a}-\bar{b}\bar{c}n+\bar{b}m+q)-h_2(n,m)\bar{\beta}\left(-q+\frac{1}{2},n-\xi \frac{m}{c} \right)\\
        &-\bar{h}_2(n,m)\bar{\kappa}\left(-q+\frac{1}{2},\frac{c}{k}n-\xi \frac{m}{k} \right)+\bar{\beta}\left(-q+\frac{1}{2},n\right)(\bar{a}+\bar{b}m-\bar{c}n+q)\\
        &\hspace{5cm}-\bar{\beta}\left(-m-q+\frac{1}{2},n\right)(\bar{a}+\bar{b}m+q)=0\ ,
    \end{aligned}
    \end{equation}
    and
    \begin{equation}\label{eq:RLbarTJacobi2}
        h_2(n,m)\beta\left(p-\frac{1}{2}, n-\xi \frac{m}{c}\right)+\bar{h}_2(n,m)\kappa \left(p-\frac{1}{2},\frac{c}{k}n-\xi \frac{m}{k}\right)=0\ .
    \end{equation}
    
    \item Ansatz \eqref{eq:RbarL-ansatz} along with \eqref{eq:LQRbarJacobi} tells us that we must have
    \begin{eqnarray}
    u_3&=&\frac{k}{c}n+\frac{m}{c} \equiv kcn+cm \nonumber\\
    v_3&=&n+\frac{m}{k} \equiv n+km\ , \nonumber
    \end{eqnarray}
    while $w_3$ identically vanishes or $t_3=kn+m$. The Jaocib identity \eqref{eq:LTRbarJacobi} implies
    \begin{equation}\label{eq:LQRbarJacobisimp}
        \begin{aligned}
             &-w_3(n,m)(a+bkn+bm+p)-h_3(n,m)\beta\left(p-\frac{1}{2},kcn+cm\right)\\
             &-\bar{h}_3(n,m)\kappa\left(p-\frac{1}{2},n+km\right)+{\kappa}\left(p-\frac{1}{2},n\right)(a+bm+kn+p)\\
             &\hspace*{4cm}-\kappa\left(m+p-\frac{1}{2},n\right)(a+bm+p)=0\ ,
        \end{aligned}
    \end{equation}
    and
    \begin{equation}
        \begin{aligned}
             h_3(n,m)\bar{\beta}\left(-q+\frac{1}{2},kcn+cm\right)+\bar{h}_3(n,m)\bar{\kappa}\left(-q+\frac{1}{2},n+km\right)=0\ .
        \end{aligned}
    \end{equation}
    \item Ansatz \eqref{eq:RbarLbar-ansatz} along with \eqref{eq:LbarQbarRbarJacobi} tells us,
    \begin{eqnarray}
    u_4&=&\frac{k}{c}n-\xi \frac{m}{c} \equiv kcn-\xi cm \nonumber\\
    v_4&=&n-\xi\frac{m}{k} \equiv n-\xi km\ , \nonumber
    \end{eqnarray}
    along with $t_4=-\bar{k}n+m$ provided if $w_4$ is non-zero. The Jacobi identity \eqref{eq:LbarTRbarJacobi} gives
    \begin{equation}\label{eq:LbarQbarRbarJacobisimp}
        \begin{aligned}
             &-w_4(n,m)(\bar{a}+\bar{b}m-\xi k\bar{b}n +q)-h_4(n,m)\bar{\beta}\left(-q+\frac{1}{2},kcn-\xi cm\right)\\
             &-\bar{h}_4(n,m)\bar{\kappa}\left(-q+\frac{1}{2},n-\xi km\right)+\bar{\kappa}\left(-q+\frac{1}{2},n\right)(\bar{a}+\bar{b}m-\bar{k}n+q)\\
             &\hspace*{4cm}-\bar{\kappa}\left(-m-q+\frac{1}{2},n\right)(\bar{a}+\bar{b}m+q)=0\ ,
        \end{aligned}
    \end{equation}
    and
    \begin{equation}
        h_4(n,m)\beta\left(p-\frac{1}{2},kcn-\xi cm\right)+\bar{h}_4(n,m)\kappa\left(p-\frac{1}{2},n-\xi km\right)=0\ .
    \end{equation}
\end{itemize}
Thus we see that simply imposing Jacobi identities in a systematic manner we have obtained the following simplified algebra involving the $R$-supercurrents
\begin{equation}
    \begin{aligned}
&[Q_r^i,R_n]=\beta(r,n) Q^i_{cn+r}\ ,\ [\bar{Q}_r^i,R_n]=\bar{\beta}(r,n) \bar{Q}^i_{\xi {c}n+r}\ ,\\
&{[Q_r^i, \bar{R}_n]}=\kappa(r,n)Q^i_{kn+r}\ ,\ [\bar{Q}_r^i, \bar{R}_n]=\bar{\kappa}(r,n)\bar{Q}^i_{\xi{k}n+r}\ ,\\
&{[T_{p,q},R_n]}=\bar{\beta}\left(-q\half,n\right)T_{p,-\xi{c}n+q}+\beta \left(p\mhalf,n\right)T_{cn+p,q}\ ,\\
&{[T_{p,q},\bar{R}_n]}=\bar{\kappa}\left(-q\half,n\right)T_{p,-\xi{k}n+q}+\kappa\left(p\mhalf,n\right)T_{kn+p,q}\ ,\\
&{[R_n, \mathcal{L}_m]}=w_1(n,m)\mathcal{L}_{cn+m}+h_1(n,m)R_{n+cm}+\bar{h}_1(n,m)\bar{R}_{kcn+km}\ ,\\
 &[R_n, \bar{\mathcal{L}}_m]=w_2(n,m)\bar{\mathcal{L}}_{-\xi cn+m}+h_2(n,m)R_{n-\xi cm}+\bar{h}_2(n,m)\bar{R}_{kcn -\xi km}\ ,\\
&[\bar{R}_n, \mathcal{L}_m]=w_3(n,m)\mathcal{L}_{kn+m}+h_3(n,m)R_{kcn+cm}+\bar{h}_3(n,m)\bar{R}_{n+km}\ ,\\
&[\bar{R}_n, \bar{\mathcal{L}}_m]=w_4(n,m)\bar{\mathcal{L}}_{-\xi kn+m}+h_4(n,m)R_{kcn-\xi cm}+\bar{h}_4(n,m)\bar{R}_{n-\xi km} \ .
\end{aligned}
\end{equation}
There is of course a number of equations constraining the form of the structure constants appearing in the above algebra. We will come back to analyse them subsequently.

Using the above, we further list down the following Jacobi identities.
\begin{enumerate}
    \item \textbf{Jacobi identity for $R_m, R_n$ and $\mathcal{L}_p$ leads to}
    \begin{equation}
    \label{eq:RRLJacobi}
        \begin{aligned}
        & w_1(n,p)w_1(m,cn+p)= w_1(m,p)w_1(n,cm+p)\ ,\\
        & h_1(m,cn+p)w_1(n,p)=  h_1(n,cm+p)w_1(m,p)\ ,\\
        & \bar{h}_1(m,cn+p)w_1(n,p)= \bar{h}_1(n,cm+p)w_1(m,p)\ .
    \end{aligned}
    \end{equation}
    \item \textbf{Jacobi identity for $R_m, R_n$ and $\bar{\mathcal{L}}_p$ leads to}
    \begin{equation}
    \label{eq:RRLbarJacobi}
        \begin{aligned}
        &w_2(n,p)w_2(m,-\xi cn+p)=w_2(m,p)w_2(n,-\xi cm+p)\ ,\\
        &h_2(m,-\xi cn+p)w_2(n,p)=  h_2(n,-\xi cm+p)w_2(m,p)\ ,\\
        &\bar{h}_2(m,-\xi cn+p)w_2(n,p)= \bar{h}_2(n,-\xi cm+p)w_2(m,p)\ .
        \end{aligned}
    \end{equation}
    \item \textbf{Jacobi identity for $\bar{R}_m, \bar{R}_n$ and ${\mathcal{L}}_p$ leads to}
    \begin{equation}
    \label{eq:RbarRbarLJacobi}
        \begin{aligned}
             &w_3(n,p)w_3(m,kn+p)=w_3(m,p)w_3(n,km+p)\ ,\\
             &h_3(m,kn+p)w_3(n,p)=  h_3(n,km+p)w_3(m,p)\ ,\\
             &\bar{h}_3(m,kn+p)w_3(n,p)= \bar{h}_3(n,km+p)w_3(m,p)\ .
        \end{aligned}
    \end{equation}
    \item \textbf{Jacobi identity for $\bar{R}_m, \bar{R}_n$ and $\bar{\mathcal{L}}_p$ leads to}
    \begin{equation}
    \label{eq:RbarRbarLbarJacobi}
        \begin{aligned}
             &w_4(n,p)w_4(m,-\xi kn+p)=w_4(m,p)w_4(n,-\xi km+p)\ ,\\
             &h_4(m,-\xi kn+p)w_4(n,p)=  h_4(n,-\xi km+p)w_4(m,p)\ ,\\
             &\bar{h}_4(m,-\xi kn+p)w_4(n,p)= \bar{h}_4(n,-\xi km+p)w_4(m,p)\ .
        \end{aligned}
    \end{equation}
    \item \textbf{Jacobi identity for ${R}_m, \bar{R}_n$ and ${\mathcal{L}}_p$ leads to}
    \begin{equation}
        \begin{aligned}
             &w_3(n,p)w_1(m,kn+p)=w_1(m,p)w_3(n,cm+p)\ ,\\
             &h_1(m,kn+p)w_3(n,p)=  h_3(n,cm+p)w_1(m,p)\ ,\\
             &\bar{h}_1(m,kn+p)w_3(n,p)= \bar{h}_3(n,cm+p)w_1(m,p)\ .
        \end{aligned}
    \end{equation}
    \item \textbf{Jacobi identity for ${R}_m, \bar{R}_n$ and $\bar{\mathcal{L}}_p$ leads to}
    \begin{equation}
    \label{eq:RRbarLbarJacobi}
        \begin{aligned}
             &w_4(n,p)w_2(m,-\xi kn+p)=w_2(m,p)w_4(n,-\xi cm+p)\ ,\\
             &h_2(m,-\xi kn+p)w_4(n,p)=  h_4(n,-\xi cm+p)w_2(m,p)\ ,\\
             &\bar{h}_2(m,-\xi kn+p)w_4(n,p)=  \bar{h}_4(n,-\xi cm+p)w_2(m,p)\ .
        \end{aligned}
    \end{equation}
\end{enumerate}

Finally, we have another family of Jacobi identities involving two superrotation generator and one $R$-supercurrent generator. The Jacobi identities for $\mathcal{L}_m, \bar{\mathcal{L}}_n, R_n$ and $\mathcal{L}_m, \bar{\mathcal{L}}_n, \bar{R}_n$ are trivially satisfied. We list down the equations that follow from the non-trivial Jacobi identities systematically
\begin{enumerate}
    \item \textbf{Jacobi identity for $\mathcal{L}_m, \mathcal{L}_n$ and $R_p$ implies}
    \begin{equation}
    \label{eq:LLRJacobi1}
        \begin{aligned}
            &(m-n)w_1(p,m+n)+w_1(p,m)(n-cp-m)- w_1(p,n)(m-cp-n)\\
            &+h_1(p,n)w_1(p+cn,m)-h_1(p,m)w_1(p+cm,n)\\
            &+\bar{h}_1(p,n)w_3(kcp+kn,m)-\bar{h}_1(p,m)w_3(kcp+km,n)=0\ ,
        \end{aligned}
    \end{equation}
    \begin{equation}
    \begin{aligned}
        (m-n)h_1(p, m+n)+& h_1(p,n)h_1(p+cn,m)-h_1(p,m)h_1(p+cm,n)=\\
        &\bar{h}_1(p,m)h_3(kcp+km,n)-\bar{h}_1(p,n)h_3(kcp+kn,m)\ ,
    \end{aligned}
    \end{equation}
    \begin{equation}
        \begin{aligned}
             (m-n)\bar{h}_1(p, m+n) +& h_1(p,n)\bar{h}_1(p+cn,m)-h_1(p,m)\bar{h}_1(p+cm,n)=\\
             &\bar{h}_1(p,m)\bar{h}_3(kcp+km,n)-\bar{h}_1(p,n)\bar{h}_3(kcp+kn,m)\ .
        \end{aligned}
    \end{equation}
    \item \textbf{Jacobi identity for $\mathcal{L}_m, \mathcal{L}_n$ and $\bar{R}_p$ implies}
    \begin{equation}
    \label{eq:LLRbarJacobi1}
        \begin{aligned}
             &(m-n)w_3(p,m+n)+w_3(p,m)(n-kp-m)- w_3(p,n)(m-kp-n)\\
             &+h_3(p,n)w_1(kcp+cn,m)-h_3(p,m)w_1(kcp+cm,n)\\
             &+\bar{h}_3(p,n)w_3(p+kn,m)-\bar{h}_3(p,m)w_3(p+km,n)=0\ ,
        \end{aligned}
    \end{equation}
    \begin{equation}
        \begin{aligned}
            (m-n)h_3(p,m+n)+& h_3(p,n)h_1(kcp+cn,m)-h_3(p,m)h_1(kcp+cm,n)=\\
            &\bar{h}_3(p,m)h_3(p+km,n)-\bar{h}_3(p,n)h_3(p+kn,m)\ , 
        \end{aligned}
    \end{equation}
    \begin{equation}
        \begin{aligned}
             (m-n)\bar{h}_3(p,m+n)+& h_3(p,n)\bar{h}_1(kcp+cn,m)-h_3(p,m)\bar{h}_1(kcp+cm,n)=\\
             &\bar{h}_3(p,m)\bar{h}_3(p+km,n)-\bar{h}_3(p,n)\bar{h}_3(p+kn,m)\ .
        \end{aligned}
    \end{equation}
   
    \item \textbf{Jacobi identity for $\bar{\mathcal{L}}_m, \bar{\mathcal{L}}_n$ and ${R}_p$ implies}
    \begin{equation}
    \label{eq:LbarLbarRJacobi1}
        \begin{aligned}
             &(m-n)w_2(p,m+n)+w_2(p,m)(n+\xi cp-m)- w_2(p,n)(m+\xi cp-n)\\
             &+h_2(p,n)w_2(p-\xi cn,m)-h_2(p,m)w_2(p-\xi cm,n)\\
             &+\bar{h}_2(p,n)w_4(kcp- \xi kn,m)-\bar{h}_2(p,m)w_4(kcp-\xi km,n)=0\ ,
        \end{aligned}
    \end{equation}
    \begin{equation}
        \begin{aligned}
             (m-n)h_2(p,m+n)+&  h_2(p,n)h_2(p-\xi cn,m)-h_2(p,m)h_2(p-\xi cm,n)=\\
             &\bar{h}_2(p,m)h_4(kc p-\xi km,n)-\bar{h}_2(p,n)h_4(kcp-\xi kn,m)\ ,
        \end{aligned}
    \end{equation}
    \begin{equation}
        \begin{aligned}
             (m-n)\bar{h}_2(p,m+n)+& h_2(p,n)\bar{h}_2(p-\xi cn,m)-h_2(p,m)\bar{h}_2(p-\xi cm,n)=\\
             &\bar{h}_2(p,m)\bar{h}_4(kcp- \xi km,n)-\bar{h}_2(p,n)\bar{h}_4(kcp-\xi kn,m)\ .
        \end{aligned}
    \end{equation}
    \item \textbf{Jacobi identity for $\bar{\mathcal{L}}_m, \bar{\mathcal{L}}_n$ and $\bar{R}_p$ implies}
    \begin{equation}
    \label{eq:LbarLbarRbarJacobi1}
        \begin{aligned}
            &(m-n)w_4(p,m+n)+w_4(p,m)(n-\xi kp-m)- w_4(p,n)(m-\xi kp-n)\\
            &+h_4(p,n)w_2(kcp-\xi cn,m)-h_4(p,m)w_2(kcp-\xi cm,n)\\
            &+\bar{h}_4(p,n)w_4(p- \xi kn,m)-\bar{h}_4(p,m)w_4(p-\xi km,n)=0\ ,
        \end{aligned}
    \end{equation}
    \begin{equation}
        \begin{aligned}
             (m-n)h_4(p,m+n)+&  h_4(p,n)h_2(kcp-\xi cn,m)-h_4(p,m)h_2(kcp-\xi cm,n)=\\
             &\bar{h}_4(p,m)h_4(p-\xi km,n)-\bar{h}_4(p,n)h_4(p-\xi kn,m)\ ,
        \end{aligned}
    \end{equation}
    \begin{equation}
        \begin{aligned}
             (m-n)\bar{h}_4(p,m+n)+& h_4(p,n)\bar{h}_2(kcp-\xi cn,m)-h_4(p,m)\bar{h}_2(kcp-\xi cm,n)=\\
             &\bar{h}_4(p,m)\bar{h}_4(p- \xi km,n)-\bar{h}_4(p,n)\bar{h}_4(p-\xi kn,m)\ .
        \end{aligned}
    \end{equation}
\end{enumerate}

\subsection{Algebra with linear structure constants} 
In our analysis so far, the structure constants have been kept arbitrary. However
the above equations are severely constraining in determining the form of the structure constants appearing above and solving them generally seems quite difficult. Hence, at this point, we make two powerful simplifying assumptions, which will somewhat reduce the complexity of the equations above. The assumptions are 
\begin{itemize}
    \item The structure constants appearing in the proposed algebra are linear in its arguments.
    \item The global sub-sector of $W(-1/2,-1/2;-1/2,-1/2)$ algebra must coincide with the  $R$-extended $\mathcal{N}=2$ super-Poincar\"{e} algebra. 
\end{itemize}
This leads us to propose an ansatz of the form
\begin{equation}\label{wi}
    \mu_i(n,p)=\omega_{i0}+\omega_{i1}n+\omega_{i2}p\ \quad \text{(for $i=1,2,3,4)$}\ ,
\end{equation}
where $\mu_i$ denotes the structure constants $w(n,p), h(n,p)$ or $\bar{h}(n,p)$. Since, for the global subsector we must have $[R_0, \mathcal{L}_m]=[\bar{R}_0,\mathcal{L}_m]=[R_0,\bar{\mathcal{L}}_m]=[\bar{R}_0,\bar{\mathcal{L}}_m]=0$, it implies that $\mu_i(0,0)=\mu_i(0,1)=\mu_i(0,-1)=0$. This along with the proposed ansatz leads us to conclude that these structure constants must be of the form
\begin{equation}\label{eq:whhbarform}
    \mu_i(n,p)=\omega n\ .
\end{equation}
where $\omega$ is an arbitrary constant. We also note that the first equation of each of the set \eqref{eq:RRLJacobi}, \eqref{eq:RRLbarJacobi}, \eqref{eq:RbarRbarLJacobi}, \eqref{eq:RbarRbarLbarJacobi} now becomes trivial.

The global sector of the $R$-supercurrent i.e. $R_0$ and $\bar{R}_0$ is supposed to commute with the generators of the Poincar\`{e} algebra $M_{\mu \nu}, P_{\mu}$ while, its commutator with the SUSY generators $Q^i_{\pm \frac{1}{2}}, \bar{Q}^i_{\pm \frac{1}{2}}$ must be nonzero. Specifically, imposing this constraint on the translation generators $P_{\mu}$, and using the identification \eqref{eq:TP0-reln}-\eqref{eq:TP3-reln}, we get,
\begin{equation}
    [T_{0,0},R_0]=[T_{0,1},R_0]=[T_{1,0},R_0]=[T_{1,1},R_0]=0\ .
\end{equation}
An identical set of relations is true for $\bar{R}_0$. The above implies that the structure constants are related by
\begin{equation}
\label{eq:betaglobal}    \bar{\beta}\left(\half,0\right)=\bar{\beta}\left(\mhalf,0\right)=-{\beta}\left(\half,0\right)=-\beta\left(\mhalf,0\right) \neq 0\ .
\end{equation}
Now, assuming that both $\beta(r,n)$ and $\bar{\beta}(r,n)$ are linear in its argument, the above relation immediately implies that they must be of the form
\begin{equation}
\label{eq:betastrconstant}
    \beta(r,n)=\beta_1 n+\beta_0 \quad \text{and} \quad \bar{\beta}(r,n)=\bar{\beta}_1 n-{\beta}_0\ ,
\end{equation}
where $\beta_1, \beta_2, \bar{\beta}_1$ and $\bar{\beta}_2$ are constants. One can repeat the same exercise with the structure constants $\kappa(r,n)$ and $\bar{\kappa}(r,n)$ to see that they are also independent of the first index $r$ and hence can be written as
\begin{equation}
\label{eq:kappastrconstant}
    \kappa(r,n)= \kappa_1 n+\kappa_0 \quad \text{and} \quad \bar{\kappa}(r,n)=\bar{\kappa}_1 n-{\kappa}_0\ .
\end{equation}
One immediate consequence of the above is that $\eqref{eq:RRbarQJacobi}$ and $\eqref{eq:RRbarQbarJacobi}$ is now trivially satisfied. With all of this conclusions, we have a more simplified algebra, given by,
\begin{equation}
    \begin{aligned}
&[Q_r^i,R_n]=\beta(n) Q^i_{cn+r}\ ,\ [\bar{Q}_r^i,R_n]=\bar{\beta}(n) \bar{Q}^i_{\xi {c}n+r}\ ,\\
&{[Q_r^i, \bar{R}_n]}=\kappa(n)Q^i_{kn+r}\ ,\ [\bar{Q}_r^i, \bar{R}_n]=\bar{\kappa}(n)\bar{Q}^i_{\xi{k}n+r}\ ,\\
&{[T_{p,q},R_n]}=\bar{\beta}\left(n\right)T_{p,-\xi{c}n+q}+\beta \left(n\right)T_{cn+p,q}\ ,\\
&{[T_{p,q},\bar{R}_n]}=\bar{\kappa}\left(n\right)T_{p,-\xi{k}n+q}+\kappa\left(n\right)T_{kn+p,q}\ ,\\
&{[R_n, \mathcal{L}_m]}=w_1(n)\mathcal{L}_{cn+m}+ h_1(n)R_{n+cm}+\bar{h}_1(n)\bar{R}_{kcn+km}\ ,\\
 &[R_n, \bar{\mathcal{L}}_m]= w_2(n)\bar{\mathcal{L}}_{-\xi cn+m}+h_2(n)R_{n-\xi cm}+\bar{h}_2(n)\bar{R}_{kcn -\xi km}\ ,\\
&[\bar{R}_n, \mathcal{L}_m]= w_3(n)\mathcal{L}_{kn+m}+h_3(n)R_{kcn+cm}+\bar{h}_3(n)\bar{R}_{n+km}\ ,\\
&[\bar{R}_n, \bar{\mathcal{L}}_m]= w_4(n)\bar{\mathcal{L}}_{-\xi kn+m}+h_4(n)R_{kcn-\xi cm}+\bar{h}_4(n)\bar{R}_{n-\xi km} \ .
\end{aligned}
\end{equation}
Now, we will focus on \eqref{eq:RLTJacobiIdentity} specifically which yields two equations given by
\begin{eqnarray}
\label{eq:LTRJacobifinalsimplified1}
w_1(n)(a+bcn+bm+p)-h_1(n)\beta\left(n+cm\right)-\bar{h}_1(n)\kappa\left(ckn+km\right)+cn\beta \left(n\right)=0\ ,\\
\label{eq:LTRJacobifinalsimplified2}
-h_1(n)\bar{\beta}\left(n+cm\right)-\bar{h}_1(n)\bar{\kappa}\left(ckn+km \right)=0\ .
\end{eqnarray}
The first equation written above must be true for arbitrary integral values of $m,n$ and $p$. However, it contains a term of the form $w_1(n)p$ which must vanish, implying that $w_1(n)=0$ identically. A similar argument can be applied to the Jacobi identities \eqref{eq:RLbarTJacobisimp}, \eqref{eq:LQRbarJacobisimp} and \eqref{eq:LbarQbarRbarJacobisimp} to reach the conclusion that $w_2=w_3=w_4=0$ too. Plugging in the linear forms of $h_1(n),\bar{h}_1(n)$ and $\beta(n)$ in the equations \eqref{eq:LTRJacobifinalsimplified1} and \eqref{eq:LTRJacobifinalsimplified2}, we get,
\begin{equation}
    \begin{aligned}
         c\beta_1 -\omega_1 \beta_1 -ck \kappa_1 \bar{\omega}_1&=0\ ,\\
         -c \omega_1 \beta_1 -k \kappa_1 \bar{\omega}_1 &=0\ ,\\
         c\beta_0 -\omega_1 \beta_0 -\kappa_0 \bar{\omega}_1 &=0\ .
    \end{aligned}
\end{equation}
The above set of equations imply $\beta_1 =0$. Applying the exact same argument to \eqref{eq:RLbarTJacobisimp}, \eqref{eq:LQRbarJacobisimp} and \eqref{eq:LbarQbarRbarJacobisimp}, yields, $\bar{\beta}_1=\kappa_1=\bar{\kappa}_1=0$. Essentially, this shows irrespective of the arguments, $\beta, \bar{\beta}, \kappa$ and $\bar{\kappa}$ are constants and related as $\beta=-\bar{\beta}$ and $\kappa=-\bar{\kappa}$. However, \eqref{eq:LTRJacobifinalsimplified1} and \eqref{eq:LTRJacobifinalsimplified2} reduces to simply
\begin{equation}
    -\omega_1 \beta_0 -\bar{\omega}_1 \kappa_0 +c\beta_0 =0 \quad \text{and} \quad \omega_1 \beta_0 +\bar{\omega}_1 \kappa_0 =0\ ,
\end{equation}
which leads us to conclude that the structure constant $\beta$ must be identically zero which is clearly in contradiction with \eqref{eq:betaglobal}. \eqref{eq:RLbarTJacobisimp}, \eqref{eq:LQRbarJacobisimp}. Similarly \eqref{eq:LbarQbarRbarJacobisimp} leads to the conclusion that $\bar{\beta}, \kappa$ and $\bar{\kappa}$ must also be vanishing. Thus we essentially find that an infinite dimensional extension of $R$-charges in $\mathcal{N}=2 \, \BMSf$ algebra is impossible with linear structure constants.
Therefore we conclude that a generic $\mathcal{N}=2$ \texorpdfstring{$W(a,b;\bar{a},\bar{b})$ algebra}{W(a,b;bar{a},bar{b})} \eqref{eq:susyW-4d} of section \ref{sec4} can not have infinite $R-$extension with linear structure constants . 

\subsection{Algebra with non-linear structure constants} 
We now relax the criteria of the linearity of structure constants and assume further that they can be at best quadratic in the arguments. Continuing to refer to the structure constants $w_i, h_i$ and $\bar{h}_i$ as $\mu_i$, we write down an ansatz of the form
\begin{equation}
    \mu_i(n,p)=\omega_{i0}+\omega_{i1}n +\omega_{i2}p+\omega_{i3}np+\omega_{i4}n^2+\omega_{i5}p^2\ .
\end{equation}
Since, $\mu_i(0,p)$ must vanish for $p=0, \pm 1$, we can write
\begin{equation}
    \mu_i(n,p)=n(\omega_{i1}+\omega_{i3}p+\omega_{i4} n)\ .
\end{equation}
Demanding \eqref{eq:betaglobal}, we see that $\beta$ and $\kappa$ (and similarly $\bar{\beta}$ and $\bar{\kappa}$) must take the form
\begin{eqnarray}
    \beta(r,n)&=&\beta_0 +\beta_2 n +\beta_3 r^2 +\beta_4 n^2 +\beta_5 rn\ ,\\
    \kappa(r,n)&=&\kappa_0 +\kappa_2 n +\kappa_3 r^2 +\kappa_4 n^2 +\kappa_5 rn
\end{eqnarray}
Unlike the linear case, \eqref{eq:RRbarQJacobi} and \eqref{eq:RRbarQbarJacobi} are not trivially satisfied now. Focussing specifically on \eqref{eq:RRbarQbarJacobi}, we recover the following equations
\begin{equation}
\begin{aligned}
    &\beta_0 \kappa_3=\beta_3 \kappa_0= \beta_2 \kappa_3 =\beta_3 \kappa_2 =\beta_3\kappa_4=\beta_4 \kappa_3=0\ ,\\
    &\beta_5 \kappa_2 = \beta_2 \kappa_5=\beta_5 \kappa_4 =\beta_4 \kappa_5=\beta_3 \kappa_3 =\beta_5 \kappa_5=0\ ,\\
    &k\beta_5 \kappa_3 = c \beta_3 \kappa_5\ ,\ k\beta_5 \kappa_0= c\beta_0 \kappa_5\ ,\\ 
    &k\beta_3 \kappa_3 +2 \beta_3 \kappa_5=c\beta_3 \kappa_3 +2 \beta_5 \kappa_3=c \beta_5\kappa_3 +2\beta_4 \kappa_3= k \beta_3 \kappa_5 + 2\beta_3 \kappa_4=0\ .
\end{aligned}
\end{equation}
We will get an analogous set of equations following from \eqref{eq:RRbarQbarJacobi}. The above set of equations implies 
\begin{equation}
    \beta_3=\kappa_3=\beta_5=\kappa_5=0\ ,
\end{equation}
thus making the structure constants $\beta(r,n)$ (and $\bar{\beta}(r,n)$) and $\kappa(r,n)$ (and $\bar{\kappa}(r,n)$) independent of $r$. Like before, \eqref{eq:betaglobal} implies
\begin{equation}
    \beta_0=-\bar{\beta}_0 \quad (\text{and similarly}\ \kappa_0=-\bar{\kappa}_0)
\end{equation}
Now, armed with the above results from \eqref{eq:RLTJacobiIdentity} we see we again recover equations very close to that of \eqref{eq:LTRJacobifinalsimplified1} and \eqref{eq:LTRJacobifinalsimplified2} which again leads us to conclude $w_i=0$ identically. Thus, in this case, the two equations following from \eqref{eq:RLTJacobiIdentity} simplifies to
\begin{eqnarray}
\label{eq:nlsimplified1}
-h_1(n,m)\beta(n+cm) -\bar{h}_1(n,m)\kappa(kcn+km)+cn\beta (n)&=&0\ ,\\
\label{eq:nlsimplified2}
-h_1(n,m)\bar{\beta}(n+cm)-\bar{h}_1(n,m)\bar{\kappa}(kcn+km)&=&0\ .
\end{eqnarray}
Plugging in the forms of the ansatze for $h_1(n,m), \bar{h}_1(n,m), \beta(n)$ and $\kappa(n)$, in \eqref{eq:nlsimplified1} and equating the various coefficients of gives
\begin{eqnarray}
\label{eq:nlprefactor11}
c\beta_0-\omega_{1}\beta_0-\bar{\omega}_1\kappa_0&=&0\ ,\\
\omega_3\beta_0+c\omega_1\beta_2+\bar{\omega}_3\kappa_0+k\bar{\omega}_1\kappa_2&=&0\ ,\\
\omega_4 \beta_0-c\beta_2+\omega_1\beta_2+\bar{\omega}_4\kappa_0+ck\bar{\omega}_1\kappa_2&=&0\ ,\\
\omega_4\beta_2-c\beta_4+\omega_1\beta_4+ck\bar{\omega}_4\kappa_2+\bar{\omega}_1\kappa_4&=&0\ ,\\
c\omega_3\beta_2+\omega_1\beta_4+k\bar{\omega}_3\kappa_2+\bar{\omega}_1\kappa_4&=&0\ ,\\
\omega_3\beta_2+c\omega_4\beta_2+2c\omega_1\beta_4+ck\bar{\omega}_3\kappa_2+k\bar{\omega}_4\kappa_2+2c\bar{\omega}_1\kappa_4&=&0\ ,\\
\omega_4\beta_4+\bar{\omega}_4\kappa_4&=&0\ ,\\
\omega_3\beta_4+\bar{\omega}_3\kappa_4&=&0\ ,\\
2c\omega_3\beta_4+\omega_4\beta_4+2c\bar{\omega}_3\kappa_4+\bar{\omega}_4\kappa_4&=&0\ ,\\
\omega_3\beta_4+2c\omega_4\beta_4+\bar{\omega}_3\kappa_4+2c\bar{\omega}_4\kappa_4&=&0\ .
\end{eqnarray}
The same exercise with \eqref{eq:nlsimplified2} gives the following equations
\begin{eqnarray}
\label{eq:nlprefactor21}
\omega_1 \beta_0 + \bar{\omega}_1 \kappa_0&=&0\ ,\\
\omega_4 \beta_0 -\omega_1 \bar{\beta}_1 + \bar{\omega}_4 \kappa_0 -ck \bar{\omega}_1 \bar{\kappa}_2&=&0\ ,\\
\omega_3 \beta_0 -c \omega_1 \bar{\beta}_1+\bar{\omega}_3 \kappa_0-k \bar{\omega}_1 \bar{\kappa}_2&=&0\ ,\\
\omega_1 \bar{\beta}_4+\omega_4 \bar{\beta}_1+ck \bar{\omega}_4 \bar{\kappa}_2+\bar{\omega}_1 \bar{\kappa}_4&=&0\ ,\\
\omega_1 \bar{\beta}_4+c \omega_3 \bar{\beta}_1 +k \bar{\omega}_3 \bar{\kappa}_2 +\bar{\omega}_1 \bar{\kappa}_4&=&0\ ,\\
2c \omega_1 \bar{\beta}_4 +\omega_3 \bar{\beta}_1 +c \omega_4 \bar{\beta}_1+ck \bar{\omega}_3 \bar{\kappa}_2+ k \bar{\omega}_4 \bar{\kappa}_2 +2c \bar{\omega}_1 \bar{\kappa}_4&=&0\ ,\\
\omega_4 \bar{\beta}_4+ \bar{\omega}_4 \bar{\kappa}_4&=&0\ ,\\
\omega_3 \bar{\beta}_4+ \bar{\omega}_3 \bar{\kappa}_4&=&0\ ,\\
2c\omega_3 \bar{\beta}_4 +\omega_4 \bar{\beta}_4+2c \bar{\omega}_3 \bar{\kappa}_4+\bar{\omega}_4 \bar{\kappa}_4&=&0\ ,\\
\omega_3 \bar{\beta}_4+2c \omega_4 \bar{\beta}_4+\bar{\omega}_3 \bar{\kappa}_4+2c \bar{\omega}_4 \bar{\kappa}_4&=&0\ .
\end{eqnarray}

The above system of equations are highly non-trivial to solve in general. But from \eqref{eq:nlprefactor11} and \eqref{eq:nlprefactor21} we easily see that
\begin{equation}
    \beta_0=0\ ,
\end{equation}
which implies $\beta(0)=0$ and similarly we find $\bar{\beta}(0)={\kappa}(0)=\bar{\kappa}(0)=0$. This is in contradiction with the constraint from the global subalgebra \eqref{eq:betaglobal}. Thus, akin to the case of linear structure constants, even for  quadratic structure constants, an extension by $U(1)_V \times U(1)_A$ realised as infinitely many generators $R_n$ and $\bar{R}_n$ ($R$-symmetry generators) of \texorpdfstring{$W(a,b,\bar{a},\bar{b})$ algebra}{W(a,b;bar{a},bar{b})} turns out to be \emph{impossible}.

\section{Conclusion}\label{sec6}

As mentioned earlier, symmetry algebras are powerful tools which severely constrain the dynamics and vacua of gauge and gravity theories. In this work, we have concentrated on supersymmetric $W(a,b)$ and $W(a,b; \bar{a},\bar{b})$ algebras which are deformations of the asymptotic symmetry algebra of supergravity theories in three and four spacetime dimensions respectively. Earlier works \cite{Parsa:2018kys, Safari:2019zmc} has established that generic deformations of $\BMSt$ algebras involve two parameters $a$ and $b$ while generic deformations of $\BMSf$ algebras involve $a,b, \bar{a}$ and $\bar{b}$. The $a=0, b=-1$ centerless $R$-extension of supersymmetric $W(a,b)$ algebra, given by \eqref{SUSY-W(a,b)} indeed matches with earlier results of \cite{Banerjee:2019lrv} where the authors performed an asymptotic symmetry analysis to obtain the super-$\BMSt$ algebra. We also classified and wrote down possible central extension of supersymmetric, $R$-extended $W(a,b)$ algebra. We observed interesting and novel central charges appearing in the $\{Q^1_r, Q^2_s\}$ anti-commutator, denoted by $f(r,s)$ for various values of $a$ and $b$. We also find that $[J_m, \mathcal{R}_n]$ admits a quadratic central charge which has not been realized through the asymptotic symmetry analysis performed in \cite{Banerjee:2019lrv} although it seems to be present for arbitrary values of $a$ and $b$. From the analysis of \cite{Banerjee:2019lrv}, it is certain that such a central term would not arise with the usual Barnich-Compere boundary conditions. Nevertheless, the study of \cite{Afshar:2015wjm} clearly indicates that with an asymptotically Rindler like behavior will modify the asymptotic symmetry algebra with such central extensions. Thus  it remains an interesting open problem to find the importance of this central term in the context of three dimensional asymptotically flat supergravity theory in more generic contexts.
As mentioned earlier,  the $W(0,0)$ and $W(0,1)$ algebras have also appeared  as asymptotic symmetry algebras of gravity theories \cite{Compere:2013bya,Afshar:2021qvi}, so it is worth  exploring appropriate boundary/falloff conditions to obtain  supersymmetric $W(0,0)$ and $W(0,1)$ algebras as asymptotic symmetry algebras in some supergravity theory. To conclude, the analysis of the present work, being mathematically rigorous, provides new asymptotic algebras and hence opens up the possibilities of finding new boundary conditions for supergravity fields. We hope to report on these possibilities in future works. 

The construction of the $R$-extended supersymmetric  $W(a,b;\bar{a},\bar{b})$ algebra turned out to be more involved. Physically, the $R$-charge generators are supposed to rotate the global SUSY-generators which motivates us to propose an ansatz of the form \eqref{eq:BMSf-RSansatz}. To be mathematically precise, we essentially tried to extend the super-$W(a,b;\bar{a},\bar{b})$ algebra (written explicitly in \eqref{eq:susyW-4d}) by a $U(1)_V \times U(1)_A$ group where each sector is represented by infinitely many generators. One of the sectors of the $U(1)_V$ symmetry can be thought of as \emph{vectorial} $R$-symmetry while the other copy of the $U(1)_A$ symmetry can be thought of as \emph{axial} $R$-symmetry. We considered two primary guiding principles to fix the algebra
\begin{itemize}
    \item For $a=b=\bar{a}=\bar{b}=-\frac{1}{2}$, the global subalgebra must be identical to the $R$-extended super-Poincar\'{e} algebra.
    \item The indices appearing in all the proposed commutators involving the $R$-charges must be linear in its arguments.
\end{itemize}
In order to simplify our calculations, we considered linear as well as quadratic structure constants. In either cases, we realized that having infinitely many $R$-charges is in contradiction with one or more Jacobi identities that must be followed by such a graded Lie algebra. Essentially, it seems there is an obstruction in the $u(1) \times u(1)$ extension of super-$W(a,b;\bar{a},\bar{b})$ algebra--which will naturally hinder the construction of an $R$-extended super-$\BMSf$ algebra. Recent work \cite{Banerjee:2021uxe} has carried out $u(1)$ and $u(N)$ extensions of $\BMSf$ by analysing celestial amplitudes of Einstein-Maxwell and Einstein-Yang-Mills theories and have indeed obtained non-trivial asymptotic symmetry algebras at the boundary which does include infinitely many generators parametrizing the $u(1)$ or $u(N)$ symmetry. This is however, not in contradiction with our results. Our demand on the behaviour of $R$-charges i.e. it must non-trivially rotate the global SUSY generators forces us to demand \eqref{eq:betaglobal} which ensures the $[Q^i_r, R_m]$ commutator to be non-zero for the global sector. Such a constraint need not be followed for the $u(1)$ or $u(N)$ gauge groups that enter the analysis of \cite{Banerjee:2021uxe}. Relaxing \eqref{eq:betaglobal} in our current work does indeed recover the symmetry algebras derived in \cite{Banerjee:2021uxe}. Finally, the methodology of our construction by throughly analysing all possible Jacobi identities along with imposing consistency with the global subsector is quite general. It is possible to adapt this algorithm to construct $u(1)$ or $u(N)$ extension for other exotic symmetry algebras. It is worth emphasizing that although we have studied $R$-extended super-$W(a,b)$ and super-$W(a,b;\bar{a},\bar{b})$ algebras in this papers, it is only for specific values of $a$ and $b$, we are aware of physical theories of gravity or supergravity where these are realized as boundary symmetry algebras. It will be interesting to explore what kind of supergravity theories give rise to these wide range of $W$-algebras for more generic values of $a, b, \bar{a}$ and $\bar{b}$. 

We would also like to mention that the construction of the present paper is not exhaustive. Cases that have not been considered here are technically difficult ones to address and there is no other clarifications/reasoning to not consider them. It will be good to find an exhaustive construction of all possible supersymmetrization of deformed $\mathfrak{bms}$ algebras and we hope to report on them in future.

Let us conclude the paper with the importance of the study of supersymmetric extensions of the deformations of $\mathfrak{bms}$ algebras. As is well understood, $\mathfrak{bms}$ algebras are symmetries of asymptotically flat gravity theories at their null boundaries. Some of their deformations have also been realized as the symmetry algebra at the horizon of certain black hole backgrounds. In the context of three spacetime dimensions, the presence of extended supersymmetries and  internal $R-$symmetries plays crucial roles in characterizing the soft hair modes (that gives non trivial cosmological solutions) and their thermodynamics \cite{Fuentealba:2017fck, Afshar:2013vka, Barnich:2012xq, Banerjee:2018hbl}. The similar study has not been performed for four space time dimensions, where non trivial black hole and gravitational wave solutions exist. In the context of $\BMSf$, the soft hair modes contributes to Black Hole entropy, albeit they do not correspond to the entire microscopic degeneracy. The microscopic degeneracy for a class of four dimensional $\mathcal{N}=2,4,8$ supersymmetric BPS black holes are very well understood \cite{David:2006ji, David:2006yn, David:2006ru, David:2006ud, Banerjee:2008ky, Sen:2009md}. It would be interesting to understand how much of this entropy is contributed by the soft hairs. Our present results suggest that, for Black Holes appearing in $\mathcal{N}=2$ spergravity theory, where the internal R-symmetry (that only scales the supercharges) will not have any contributions to the soft hairs. The similar study for other supergravity theories with exotic internal symmetries remain an open problem to study in the future. 

\section*{Acknowledgements}
We would like to thank M. M. Sheikh-Jabbari for collaboration and useful discussions during the initial course of this work and for his comments on the manuscript. DM would like to thank IISER Bhopal for kind hospitality during the end of this project. The work of NB is partially supported by SERB ECR grant ECR/2018/001255. The work of DM is supported in part by the SERB core research grant CRG/2018/002373. Finally we thank the people of India for their generous support to fundamental research. HRS acknowledges the support of Saramadan grants ISEF/M/99407. \\

\appendix

\section{Solution to structure constants $\sigma(m,n)$ and $\kappa(m,n)$ for $W(a,b)$ algebra}
\label{app:sigma-kappasoln}

We briefly provide the details of solution for the system of equations as given in \eqref{eq:sigma-kappa-system1}-\eqref{eq:sigma-kappa-system4}. We will assume that the deformation parameter $b \neq 0$. Clearly, from the first equation of \eqref{eq:sigma-kappa-system1}, it follows $\sigma_1=0$. This reduces the effective system of equations to the following:
\begin{eqnarray}
\label{eq:sigma1zero}
\sigma_0 (\kappa_0-a)=0\ ,\ &\quad&\ \sigma_2(\kappa_2-b-1)=0\ ,\\
\sigma_0 (\kappa_2-b)=0\ ,\ &\quad&\ \sigma_2(\kappa_0-a)=0\ .
\end{eqnarray}
From the first equation of \eqref{eq:sigma1zero} we can conclude
\begin{equation}
    \text{either}\ \sigma_0=0\ \text{or}\ \kappa_0=a\ .
\end{equation}
In the former case when $\sigma_0=0$, we end up with two effective equations, namely,
\begin{equation}
    \sigma_2(\kappa_2-b-1)=\sigma_2(\kappa_0-a)=0\ .
\end{equation}
For the case when $\sigma_0=\sigma_1=0$, a non-zero structure constant $\sigma(m,n)$ implies $\sigma_2 \neq 0$ but arbitrary while we must have 
\begin{equation}
    \kappa_0=a\ \text{and}\ \kappa_2=b+1\ .
\end{equation}
For the other case, when $\kappa_0=a$, the effective system of equation is
\begin{equation}
    \sigma_0(\kappa_2-b)=\sigma_2(\kappa_2-b-1)=0\ .
\end{equation}
For a non-zero $\sigma_0$, we must have $\kappa_2=b$ and $\sigma_2=0$. These cases exhaust all possible solutions to the system of equations \eqref{eq:sigma-kappa-system1}-\eqref{eq:sigma-kappa-system4}.

\section{Solution to index structure of the fermionic supercurrent anticommutator}
\label{app:QQbarstructure}

In order to fix the index structure as per the ansatz \eqref{eq:QQ-indexansatz}, the index structure must satisfy the set of equations given by \eqref{eq:poincarealgfg} in order to be consistent with super-Poincar\'{e} algebra. We have eight equations governing six quantities, namely $f_i,g_i$ for $i=0,1,2$. Although this system is overconstrained, we can indeed find a consistent set of solutions for the above. Using the condition $f\left(\half,\half\right)=f\left(\half,\mhalf\right)=1$, we end up with $f_2=0$ while the condition $g\left(\half,\mhalf\right)=g\left(\mhalf,\mhalf\right)=1$ gives $g_1=0$. The above four conditions essentially gives us two equations
\begin{equation}
    f_0 +\frac{f_1}{2}=g_0-\frac{g_2}{2}=1\ .
\end{equation}
The two other conditions on $f(r,s)$ given by $f\left(\mhalf,\half\right)=f\left(\mhalf,\mhalf\right)=0$ results in a single equation
\begin{equation}
    f_0 -\frac{f_2}{2}=0\ ,
\end{equation}
which finally yields 
\begin{equation}
    f_0=\frac{1}{2} \quad \text{and}\quad f_1=1\ .
\end{equation}
The conditions $g\left(\half,\half\right)=g\left(\mhalf,\half\right)=0$ results in the equation
\begin{equation}
    g_0+\frac{g_2}{2}=0\ ,
\end{equation}
which finally gives 
\begin{equation}
    g_0=\frac{1}{2} \quad \text{and}\quad g_2=-1\ .
\end{equation}
Thus, we see although it seems this might be an over constrained system of equations, we can find a consistent solution set which we report in \eqref{eq:fgfullsolution}.

\bibliographystyle{JHEP}
\bibliography{References_draft}

\end{document}